Latex this file
\documentstyle[preprint,revtex]{aps}
\begin{document}
\draft
\hfill\vbox{\hbox{\bf NUHEP-TH-93-3}\hbox{Feb 1993}}\par
\thispagestyle{empty}
\begin{title}
{\bf Search for the intermediate Mass Higgs Signal at TeV $e\gamma$ colliders}
\end{title}
\author{Kingman~Cheung}
\begin{instit}
Dept. of Physics \& Astronomy, Northwestern University, Evanston,
Illinois 60208, USA\\
\end{instit}
\begin{abstract}
\nonum
\section{Abstract}
The intermediate mass Higgs (IMH) can be abundantly produced through the
process $e^-\gamma \rightarrow W^-H\nu$ at TeV $e^-\gamma$ colliders, which are
realized by the laser back-scattering method.  We search for the signature of
$W^-H \rightarrow (jj)(b\bar b)$ plus missing transverse momentum, with
and without considering the $b$-tagging.  We also analyse all the potential
backgrounds from $e^-\gamma \rightarrow W^-Z\nu,\,W^-W^+e^-,\,ZZe^-,\,
\bar t b\nu$ and $t\bar t e^-$.
With our selective acceptance cuts these backgrounds are reduced to a
manageable level.  We find that for the entire intermediate mass range
60 -- 150~GeV the Higgs discovery should be viable.
We also present detail formulas for the  helicity amplitudes of
these processes.
\end{abstract}

\newpage
\section{Introduction}
\label{intro}

The symmetry-breaking sector of the standard model (SM) is the most mysterious
part of the particle theory.  Even for  the simplest
minimal SM the Higgs boson, which is  responsible for the symmetry-breaking,
has  not yet been
found, and there is no theoretical restriction on its mass except the
unitarity implies  an upper limit of about 1~TeV on its mass.
The discovery of the Higgs boson  depends
on its mass, which determines the decay channel to search for.  For the heavy
Higgs ($m_H\agt 2m_Z$) we can use the gold-plated  channel,
 $H\rightarrow ZZ \rightarrow \ell\bar\ell\ell\bar\ell$ to identify
it at hadronic colliders\cite{ZZ}, and even the four-jet mode of
$H\rightarrow ZZ,\,WW \rightarrow
(jj)(jj)$ at $e^+e^-$ colliders\cite{ee-cheung,ee-hagi}.
The present lower bound on $m_H$ from
LEP is about 52 -- 53~GeV\cite{LEP}, which can extend up to  60~GeV in
near future. There remains a mass range of 60 to 140~GeV that the Higgs, which
predominately decays into $b\bar b$ pair,  could be
difficult to  identify because of large hadronic background at hadronic
colliders.  Recent studies
showed that we can use the rare photonic mode of IMH decaying
into $\gamma\gamma$ to search for the Higgs boson in the direct
$gg\rightarrow H$\cite{ggH} production, or in the associated
productions with $W$-boson\cite{ass-w}  or $t\bar t$ pair\cite{ass-tt}.
While  at $e^+e^-$ colliders, we can use the $e^+e^- \rightarrow ZH
\rightarrow (f\bar f) b \bar b$ to identify the IMH  upto about 90 -- 95~GeV
at LEP II \cite{janot}, and the whole intermediate mass range at
Next Linear Collider (NLC) \cite{janot}.

With the recent discussions of converting the linear $e^+e^-$ colliders into
$\gamma\gamma$ or $e\gamma$ colliders by laser back-scattering method, they
provide new physics possibilities of detecting and probing the properties of
the Higgs boson\cite{bord}.  With a 0.5 TeV
$e^+e^-$ collider in $\gamma\gamma$ mode,  the Higgs production by
photon-photon fusion via a triangular loop of heavy fermions and $W$-boson can
be used to discover a heavy Higgs boson ($m_H>2m_Z$) \cite{bord,richard,dave}.
  It was
shown in Ref.~\cite{dave} that a heavy Higgs boson of mass up to about 350~GeV
should be able to be identified in the decay mode of $H\rightarrow
ZZ\rightarrow q\bar q\ell\bar\ell$, which has a sufficiently large branching
fraction and is free from the huge $\gamma\gamma\rightarrow WW$ background.
It was also shown in Ref.~\cite{dave} that the detectability in general
decreases for a higher energy machine.
On the other hand, the
process $\gamma\gamma\rightarrow t\bar tH$\cite{ttH} is shown to be better
than the corresponding process, $e^+e^-\rightarrow t\bar tH$, in $e^+e^-$
collider for the measurement of the Yukawa top-Higgs
coupling at $\sqrt{s}=1-2$~TeV.

Another interesting Higgs production process is
$e\gamma\rightarrow WH\nu$\cite{boos,hagi} by colliding a
photon beam with an electron or positron beam.  The cross section of this
process is shown to be comparable to $e^+e^-\rightarrow \nu\bar\nu W^*W^*
\rightarrow \nu\bar\nu H$ at $\sqrt{s}=1-2$~TeV, and much larger than the
Bjorken process $e^+e^-\rightarrow ZH$ for IMH mass range.  However, the
backgrounds have not fully analysed.  The major backgrounds for IMH search
in the process $e^-\gamma\rightarrow W^-H\nu$ come from $e^-\gamma\rightarrow
W^-Z\nu,\, WWe^-$ and  $ZZe^-$, in which the boson-pair decay hadronically
into four jets in the final state.  Also there are backgrounds from $e^-\gamma
\rightarrow b\bar t\nu\rightarrow b\bar bW^- \nu$ and $e^-\gamma\rightarrow
t\bar te^-\rightarrow b\bar bWWe^-$.  With the $b$-identification
these backgrounds
can be much reduced, however, the $b$-tagging efficiency is so far uncertain.
   It is then the
purpose of this paper to investigate the feasibility of identifying the IMH
through the process $e^-\gamma\rightarrow W^-H\nu$ at TeV $e\gamma$ colliders,
with and without implementing the $b$-tagging.
The organization is as follows: we will describe the calculation of
the signal and background processes  including the photon luminosity function
in Sec.~\ref{secII}, following which  the results
will be presented in according to the case of with and without implementing
$b$-tagging
in Sec.~\ref{secIII}.  We will then summarize the conclusions and discussions
in Sec.~\ref{secIV}.  We will
also give detail formulas for the processes involved in the
Appendix~\ref{app}.

\section{Calculation Methods}
\label{secII}

\subsection{Photon luminosities}
\label{photon-lum}

We use the energy spectrum of the back-scattered photon given by
\cite{teln}
\begin{equation}
\label{lum}
F_{\gamma /e}(x) = \frac{1}{D(\xi)} \left[ 1-x +\frac{1}{1-x}
-\frac{4x}{\xi(1-x)} + \frac{4x^2}{\xi^2 (1-x)^2} \right] \,,
\end{equation}
where
\begin{equation}
\label{D_xi}
D(\xi) = (1-\frac{4}{\xi} -\frac{8}{\xi^2}) \ln(1+\xi) + \frac{1}{2} +
\frac{8}{\xi} - \frac{1}{2(1+\xi)^2}\,, \\
\end{equation}
$\xi = 4E_0\omega_0/m_e^2$,  and $\omega_0$ is the energy of
the incident laser photon.   $x=\omega/E_0$ is the fraction of the energy
of the incident positron carried by the back-scattered photon, and
the maximum value  $x_{\rm max}$ is given by
\begin{equation}
x_{\rm max} = \frac{\xi}{1+\xi}\,.
\end{equation}
The value for $\xi$ is chosen in such a way that the back-scattered photon will
 not produce the unwanted $e^+e^-$ pairs with the incident laser photon.
We choose $\xi$ to be 4.8, and so $x_{\rm max}\simeq 0.83$,
$D(\xi)\simeq 1.8$, and  $\omega_0= 1.25(0.63)$~eV for a 0.5(1) TeV $e^+e^-$
collider.
Here we have assumed that the positron and the back-scattered photon
beams are unpolarized.  We also assume that, on average, the number of
the back-scattered photons produced per positron is 1.

Besides, photon is also known to interact via its quark and gluon
constituents \cite{witten}. This is referred as a ``resolved" photon process.
The gluons and quarks are treated as  partons inside the photon with the
distribution functions $P_{i/\gamma}(x)$ to describe the probability that the
parton $i$ carries  a momentum fraction $x$.  What we need is  the gluon
distribution function to calculate some backgrounds wherever the electron-gluon
scattering also contributes in the case of electron-photon scattering.
However,  the gluon parton distribution function inside photon has
large uncertainty because limited experimental data are available.
We choose the  parameterization of Drees and Grassie (DG)~\cite{DG}
for the photon structure function,
a scale $Q^2= \hat s /4$ and $\Lambda_4$ to be 0.4~GeV for both the photon
structure function and $\alpha_s$ (evaluated at the second order).

The subprocess cross sections $\hat \sigma$ must be folded with
the luminosities to find the total  cross sections.
In case of electron-photon scattering, the cross section $\sigma$ is
\begin{equation}
\sigma(s) = \int_{x_{1min}}^{x_{max}}dx_1 \; F_{\gamma/e}(x_1) \hat
\sigma(\hat s=x_1s) \,.
\end{equation}
For the electron-gluon scattering the total cross section is given by
\begin{equation}
\sigma(s) = \int_{x_{1min}}^{x_{max}} dx_1\,\int_{x_{2min}}^1  dx_2 \;
F_{\gamma/e}(x_1) \,P_{g/\gamma}(x_2) \hat \sigma(\hat s=x_1x_2s) \,.
\end{equation}

\subsection{$e\gamma \rightarrow WH\nu$}
\label{WH}

The contributing Feynman diagrams are shown in Fig.~\ref{fey-wh}.  This
process has been calculated in details in Refs.~\cite{boos,hagi}. For
completeness detail formulas for the matrix elements are given in
Appendix~\ref{app}. We did an independent calculation that agrees with their
results.  The gluon parton inside photon does not contribute because the gluon
does not couple to either initial or final state particles.  For the Higgs
in the intermediate
mass range the signature, due to the dominate decay of $H\rightarrow b\bar b$
and the hadronic decay of $W$, will be
\begin{equation}
e^-\gamma \rightarrow W^- H\nu \rightarrow (jj) (b\bar b) \nu\,,
\end{equation}
where there are four jets plus missing energy in the final state.  Two of the
four jets are reconstructed at the $W$ mass, and the other two can be
reconstructed as a resonance peak at the Higgs mass.  For this signature the
direct backgrounds are the $W^-Z$, $W^-W^+$ and $ZZ$ productions where the $W$
and $Z$ decay hadronically into four jets, especially if the Higgs mass is
close to $W$ or $Z$ mass.  These processes will be described next.

\subsection{$e^-\gamma \rightarrow W^-Z\nu,\,W^-W^+e^-$ and $ZZe^-$}
\label{WZ}

These processes have been calculated in Ref.~\cite{me}.
There are totally 11 contributing
Feynman diagrams in the process $e^-\gamma\rightarrow W^-Z\nu$, 18 in
$e^-\gamma\rightarrow W^-W^+e^-$, and 6 in $e^-\gamma\rightarrow ZZe^-$, in
the general $R_\xi$ gauge, shown in Fig.~\ref{fey-wz}.
The $W^-Z$ and $W^-W^+$ productions are interesting by
their own in the subject of probing the triple and quartic gauge-boson
interactions \cite{me}.  The formulas for the Feynman amplitudes of these
processes  are presented here again in Appendix~\ref{app}.
The $WW$ production starts with a hugh cross section (see Fig.~\ref{cross}),
but  it was shown in Ref.~\cite{me} that a transverse
momentum $p_T(VV)$ cut on the boson pair can reduce the $W^-W^+$ background
substantially, and only moderately on the signal and $W^-Z$.  Also we will
show that the central
electron vetoing method will be very useful in further reducing the $WW$
background.
The $W^-W^+$ background is reducible if
100\% $b$-tagging is used, since we can require
$b\bar b$ pair in the final state.  The $W^-W^+$ background decaying  into $b$
and $\bar b$ is well suppressed by the Cabbibo angle.  Otherwise, if no
$b$-tagging, we have to consider all these direct backgrounds.

\subsection{$e^-\gamma \rightarrow b \bar t \nu$}
\label{bt}

The contributing Feynman diagrams are shown in Fig.~\ref{fey-bt}(a).  This
process was calculated in great details in Ref.~\cite{jikia}.  The formulas for
the matrix elements are also given in Appendix~\ref{app}.
Its cross section is of order 0.02 -- 0.1~pb for the energy range
of $\sqrt{s_{e^+e^-}}=0.5$ -- 2~TeV.  The $\bar
t$ so produced will decay 100\% into $\bar b W^-$ so that it can mimic
the signal
because there are, therefore, a $W^-$ and $b\bar b$ pair plus missing energy
due to the missing neutrino in the final state.
However, the $b\bar b$ pair in this production does  not likely form a sharp
peak but a continuum background.  The feasibility of detecting the $b\bar b$
pair resonance peak from the Higgs decay
depends on whether the Higgs peak stands significantly ({\it e.g.},
$S/\sqrt{B}>4$) above the continuum, and whether there are enough events under
the Higgs-peak.  We did an independent calculation, which agree with
the results in Ref.~\cite{jikia},  with full spin-correlation
in the subsequent decay of $\bar t\rightarrow \bar bW^-$.  Here we used a
top-quark mass $m_t=150$~GeV and a bottom-quark mass $m_b=4.5$~GeV
to illustrate.

The gluon parton inside the photon also contribute via electron-gluon
scattering.  Using the gluon distribution function described in
Sec.~\ref{photon-lum}, the contribution from the electron-gluon scattering can
be as large as 50\% at $\sqrt{s_{e^+e^-}}=2$~TeV for DG, though it is only
about 15\% at 1 TeV, and negligible at 0.5 TeV.  We will show the contribution
{} from this ``resolved" photon process in Fig.~\ref{cross} and in
our final results in the Tables,
otherwise this contribution is left out in all the other figures.

\subsection{$e^-\gamma\rightarrow t \bar t e^-$}
\label{tt}

The contributing Feynman diagrams are shown in Fig.~\ref{fey-bt}(b), and the
formulas for the helicity amplitudes are given in Appendix~\ref{app}.
This is also a potential background when the $t$ and $\bar t$ decay into $bW^+$
and $\bar b W^-$ respectively, and only one of the $W$ is detected.  We assume
that the more energetic $W$ is the one detected.  In the
final state there are therefore a $W$, $b\bar b$ pair plus missing energy due
to the other undetected $W$ and $e^-$.  Similar to the previous background,
the $b\bar b$ pair from $t\bar te^-$ will not form a sharp resonance
peak but a continuum.
 Here we included full spin-correlation in the subsequent decays of
$t$ and $\bar t$, and take $m_t=150$~GeV.  In this calculation we have
neglected  the contribution from ``resolved" photon process,
because it  needs  a rather high energy threshold for
producing $t\bar t$ pair, where the gluon luminosity function drops to a
small value.
Other background arising from the QCD production
of four jets are  $\alpha_s^2$ suppressed relative to the signal even before
imposing the constraint of $W$ or $Z$ mass on the invariant mass of jet
pairs, so these QCD backgrounds are negligible.

\section{Results}
\label{secIII}

We use the following input parameters: $\alpha_{\rm w}=1/128$,
$m_Z=91.175$~GeV and
$x_{\rm w}=0.23$, which on tree-level gives $m_W=80.0057$~GeV.
We show the total cross-sections for the signal and various backgrounds with
$m_t=150$ and $m_H=100$~GeV for the center-of-mass energies
$\sqrt{s_{e^+e^-}}=0.5-2$~TeV in Fig.~\ref{cross}.  We can see that
the cross section of $W^-Z$ is of order 0.1 -- 1~pb for the energy range shown
and the $W^-W^+$ production is of order 4 -- 20~pb.
On the other hand, the $ZZ$ production is relatively negligible, and the
signal $WH$ production is only of order 0.01 -- 0.2~pb.  The other two
backgrounds from $e^-\gamma\rightarrow \bar t b\nu$ and $t\bar te^-$
are of more or less the same size as the $WH$ signal.
As mentioned above, the $p_T(VV)$ spectrum of the boson-pair can help to
differentiate the signal and various backgrounds.  In Fig.~\ref{ptvv} we show
the dependence of the differential cross section $d\sigma/dp_T(VV)$ on the
transverse momentum of the boson-pair at $\sqrt{s}=1$ and 2~TeV.  In this
figure, we did not include any branching fractions of the bosons.  From
this figure, we can choose an acceptance cut of
\begin{equation}
\label{ptvvcut}
p_T(VV)  > \left\{\quad \begin{array}{ll}
                15\;{\rm GeV}\quad {\rm for} & \quad \sqrt{s}=1\;{\rm TeV}\\
                30\;{\rm GeV}\quad {\rm for} & \quad \sqrt{s}=2\;{\rm TeV}\,,
                  \end{array}
           \right .
\end{equation}
to reduce the $WW$ background.  We can also use the
central electron vetoing\cite{ee-cheung,ee-hagi},
{\it i.e.}, rejecting events with
electrons detected in the central region,
\begin{equation}
\label{vetoing}
E(e)  >  50\;{\rm GeV}\qquad {\rm and} \qquad |\cos\theta_e|
< \cos(0.15)\,,
\end{equation}
to further reduce backgrounds that have $e^-$ in the final state.  Totally a
factor of 10 reduction on $WW$ background is achieved
by combining  the  cuts of Eqs.~(\ref{ptvvcut}) and (\ref{vetoing}),
whereas it has almost no effect on the signal (see Table~\ref{tableI}).

Further reduction of backgrounds  can be made possible by analyzing
the direction of the outgoing boson-pair.  We define the direction of the
incoming
$e^-$ beam  as the positive $z$-axis, and so the incoming photon beam as the
negative $z$-axis.
We  select the $W$-boson in $WZ$ production, $W$ in $WH$, either
$W$ in $WW$, either $Z$ in $ZZ$, $W$ in $\bar tb\nu\rightarrow Wb\bar b\nu$,
 and the more energetic $W$ in $t\bar t e^-\rightarrow b\bar b WWe^-$, as
the boson $V_1$;  and so the $Z$-boson in $WZ$, the $H$ in $WH$,
the other $W$ in $WW$, the other $Z$ in $ZZ$, the $b\bar b$ pair in $\bar
tb\nu \rightarrow Wb\bar b\nu$ and the $b\bar b$ pair in $t\bar t \rightarrow
b\bar bWW e^-$ as the boson $V_2$.
We then show the dependence of the differential cross-section on
the cosine of the angle between
the positive $z$-axis and the direction of the boson $V_1$ and $V_2$
in Fig.~\ref{cosv1v2}(a) and (b), respectively.
We can see that the $WW$, $ZZ$, $b\bar t\nu$ and $t\bar te^-$ backgrounds
statistically have both bosons in the same hemisphere more often
than in opposite
ones, where the hemispheres  are defined as the two half spaces separated
by the plane that is perpendicular to the $z$-axis and contains the
collision point.  On the other hand, the events of $WZ$ and $WH$ tend to
have the boson-pair coming out  in opposite hemisphere.
Therefore, we can reduce backgrounds by
requiring the two bosons to come out in opposite hemisphere, {\it i.e.},
\begin{equation}
\label{cosvvcut}
\cos\theta_{V_1} \cos\theta_{V_2} <0 \,.
\end{equation}
We also show the spectrum of
$\cos\theta_{V_1} \cos\theta_{V_2}$ in Fig.~\ref{cosv1v2}(c).
Actually, we could have been requiring $\cos\theta_{V_1}<0$ and
$\cos\theta_{V_2}>0$, {\it i.e.}, $V_1$ going out in the ``negative"
($\cos\theta <0$) hemisphere and $V_2$ going out in the ``positive"
($\cos\theta >0$) hemisphere.  We expect this additional  acceptance cut
could further reduce backgrounds by a large amount (see Fig.~\ref{cosv1v2}(a)
and (b)).  However, there are uncertainties in determining which boson is $V_1$
or $V_2$ experimentally in the case of no $b$-tagging, and in the case of
the backgrounds from $ZZ$ and $WW$,
plus the situation when $m_H$ overlaps with $m_Z$ or $m_W$
then we could not determine which jet pair forms $V_1$ or $V_2$.
Therefore, we only employ the cut in Eq.~(\ref{cosvvcut})  in the angular
distributions of $V_1$ and $V_2$ so that we are safe from the above
uncertainties.
We can see from Table~\ref{tableI} that the cut of Eq.~(\ref{cosvvcut})
actually cuts more on $WW$, $ZZ$, $\bar tb\nu$ and $t\bar te^-$  than on the
$WH$ and $WZ$.
We summarize in
Table~\ref{tableI} the effectiveness of various combinations of cuts of
Eq.~(\ref{ptvvcut}), (\ref{vetoing}) and (\ref{cosvvcut}).
After all these cuts, we can
proceed to look at the invariant  mass spectrum of the two jets that comes
{} from the decay of the Higgs-boson or $V_2$.

For the following we will consider two extreme cases:
(a) with 100\% efficient $b$-tagging and (b) without $b$-tagging.  In the real
experiment the situation will be in between these two extreme cases.  For 100\%
efficient $b$-tagging the $WWe^-$ background drops because we can require a
$b\bar b$ pair in the final state and the decay of $WW$ pair into $b\bar b$
pair is strongly suppressed by the Cabbibo mixing.  Nevertheless, for the case
of no $b$-tagging we have to consider all the backgrounds.

\subsection{With 100\% efficient $b$-tagging}

Since the IMH predominately decays into $b\bar b$ pair ($\approx$ 0.8--0.9)
whereas
the $Z$ boson decay only 15\% into $b\bar b$ (but about 70\% into jets),
therefore $b$-tagging can reduce the $WZ$ and $ZZ$ backgrounds by a
factor of four to five.  Note that $e\gamma \rightarrow e^-b\bar b Z$ is
an order $\alpha_{\rm w}$ suppressed.   As mentioned above, the $WW$
background is reducible with 100\% $b$-tagging.
The invariant mass $m(b\bar b)$ spectra for the signal and various
backgrounds are shown in Fig.~\ref{mbb-yes} for $\sqrt{s_{e^+e^-}}=1$
and 2~TeV, in
which the branching fractions of $V_1 \rightarrow jj$ and $V_2 \rightarrow
b\bar b$ are included.
We use the $B(Z,W\rightarrow jj)\simeq 0.7$, $B(Z\rightarrow b\bar b)=0.15$
and the $B(H\rightarrow b\bar b)$ from Ref.~\cite{ee-cheung}.
As expected the $b\bar b$ pair from $b\bar t\nu$ and
$t\bar te^-$ productions form continuum spectra, while those from
$WZ\nu$, $ZZe^-$ and $WH\nu$ form discrete sharp peaks.  These peaks,
in collider
experiments, actually spread out due to the resolution of the detector, though
the Higgs width is very narrow for the intermediate mass range that we are
considering.  We assume the peaks spread out uniformly over a range of
$\pm 5$~GeV about the central values ($m_Z,\,m_W$ or $m_H$).
We also assume that
the Higgs peak is isolated if it is 10~GeV or more away from the $Z$-mass.
In this case, the signal $S$ is simply the cross section  under the isolated
peak, and
the background $B$ is the continuum background with $m(b\bar b)$
falling in between $m_H\pm 5$~GeV.
On the other hand, if $|m_H-m_Z|<10$~GeV the Higgs and $Z$ peaks are
overlapping.  In this case, we have to include the whole or part of the
$Z$-peak into the background $B$.  Naively, we can take a  linear fraction
of the $Z$-peak
\begin{equation}
\label{masscut}
\sigma(Z\;{\rm peak}) \times \frac{{\rm max}\left( 0,\,10\;{\rm GeV}\; -|m_H
- m_Z| \right )}  {10\;{\rm GeV}}
\end{equation}
plus the continuum in between $m_H\pm 5$~GeV as the total background $B$.

In Fig.~\ref{mbb-yes}, the continuum backgrounds from $\bar tb\nu$ and
$t\bar te^-$ are rather flat,  and far below the Higgs or $Z$ peak.
In this figure we show the Higgs-peak for $m_H=100$~GeV, which is already
slightly higher than the $Z$-peak.  So we expect when the $m_H$ goes down to 60
GeV (LEP limit) the Higgs peak will become higher because of the increase in
both $\sigma(e\gamma\rightarrow WH\nu)$ and $B(H\rightarrow b\bar b)$ as $m_H$
decreases.  Hence, we expect the discovery of the Higgs even
with $m_H \simeq m_Z$
to be viable by employing the $b$-tagging.
On the other hand, when $m_H$ increases from 100 GeV the Higgs peak will
decrease because of the decrease in both $\sigma(e\gamma\rightarrow WH\nu)$
and $B(H\rightarrow b\bar b)$; especially after $m_H=140$ GeV, the
$B(H\rightarrow b\bar b)$ drops sharply.  Fortunately, at this range $m_H\agt
100$~GeV the Higgs peak should be far away enough from the $Z$-peak, and
the continuum backgrounds are far below.  Therefore,
 the discovery of the Higgs depends
only on the actual number of events under the Higgs peak.
In Table~\ref{tableII}, we show the cross-sections in femtobarn
for the signal $S$, various  backgrounds, total background $B$
and the corresponding significance $S/\sqrt{B}$ of the signal, for various
values of $m_H$ from 60 -- 160~GeV at $\sqrt{s_{e^+e^-}}=1(2)$~TeV,
with an assumed integrated luminosity of 10~fb$^{-1}$.
We assume  a signal of 6 or more events with a significance
greater than 4 for the discovery of an isolated Higgs-peak;
whereas in the case of
overlapping Higgs-peak we  require $S \ge10$ with $S/\sqrt{B}>6$   for
discovery.  With this criterion the Higgs boson can be discovered for
$m_H=60$ to 150~GeV and marginally up to 160 GeV  (see Table~\ref{tableII})
in $b\bar b$ decay mode, providing 100\% efficient
$b$-identification.

The signal for $m_H\simeq m_Z$ is slightly larger than the background, so
with sufficient number of signal events the Higgs discovery at the $Z$-peak
should be viable without knowing exactly the absolute normalization of the
$Z$-peak.  For those Higgs-masses away from the $Z$-mass the continuum
background is so small that the actual
number of signal events,which  is the most important factor for Higgs
discovery, is large enough upto $m_H=150$~GeV.
But as $m_H$ increases from 150~GeV, the $B(H\rightarrow b\bar b)$
goes down sharply from 18\% to 3.7\% at $m_H=160$~GeV, and the number of
signal events become marginal for discovery.  Especially
 after $m_H=160$~GeV, there are too few signal events for discovery

So far we have assumed 100\% acceptance and detection efficiencies for the
jets decayed from the boson-pair.  If we take into account the overall
acceptance and detection efficiencies, say 25\% overall, the number of signal
and background events will decrease to 25\%, and the significance $S/\sqrt{B}$
will be halved.  Even that we still have sufficient number of signal events
and large enough $S/\sqrt{B}$ to cover the whole range of $m_H=60-150$~GeV,
including $m_H\simeq m_Z$.

\subsection{Without $b$-tagging}

If without $b$-tagging there are several combinations of the four jets in the
final state.  One way to select the events is to pick out those that have two
of the four jets reconstructed at the $W$-mass ($Z$-mass for $ZZ$, but it is
negligible), then take the other
two jets for considering the Higgs bosons.   We assume this reconstruction can
select the signal and the relevant background events very efficiently.
We plot the spectrum of invariant mass of the third and fourth jets in
Fig.~\ref{mbb-no}.  We can see that
\begin{description}

\item {(i)}the backgrounds from $WZ$, $WW$,  after
picking out the $W$ (either  $Z$ for $ZZ$), form discrete peaks at
either $W$ or $Z$ mass;

\item {(ii)}the background from $b\bar t\nu \rightarrow b\bar b W^-\nu$,
after picking out
the $W$, the remaining $b\bar b$ can only form a continuum;

\item {(iii)}for the $t\bar te^-\rightarrow b\bar b WWe^-$, after picking
out the more energetic  $W$, we assume that we did not pick out the
jet-pair from the other $W$.
Therefore, the other two jets (from $b\bar b$) forms a continuum
invariant-mass spectrum.  Here we assumed this procedure is valid for
our analysis,
though experimentally we might  pick out any two of the remaining four jets,
or we might have picked out more than two parton-jets when they are close to
one another;

\item{(iv)}for the signal, after picking out the $W$, the $b\bar b$ will form
a discrete peak at $m_H$.

\end{description}

The continuum backgrounds are rather flat, and far below the $W$, $Z$ and Higgs
peaks.  So when the Higgs-peak is isolated it should be able to be discovered,
provided that it has sufficient number of events under the peak.
In this figure we
show the Higgs peak for $m_H=100$~GeV, and the $Z$-peak is about four times
and the $W$-peak is eight times as high as the Higgs-peak.
The Higgs peak, for the same reason mentioned in the last subsection,
will become  higher when $m_H$ decreases, and  smaller when $m_H$ increases.
Here we also  have the cases whether the Higgs peak overlaps
with the $Z/W$-peak or the Higgs peak is isolated.  We take the same
treatment as in the last subsection for the signal
$S$ and background $B$, but here we used the branching fractions of $V_1
\rightarrow jj$ and $V_2 \rightarrow jj$, and  present  the results in
Table~\ref{tableIII}.
We assume a signal rate of 6 or more events with
significance greater than 4  for the discovery of an isolated Higgs-peak;
when the Higgs-peak
overlaps with the $W$ or $Z$-peak, absolute  normalization of the $W$ or $Z$-
peak is
important that  we require more signal events ($\agt 10$) with larger
significance ($\agt 6$) for the Higgs discovery in order to change the
absolute normalization of the $W$ or $Z$- peak by a significant  amount.
With this criterion, from Table~\ref{tableIII} we should be able to
discover the whole intermediate mass range of 60 to 150~GeV,
and marginally upto 160~GeV.

The signal for $m_H\simeq m_W$ and  $m_Z$ is about $\frac{1}{6}$ and
$\frac{1}{3}$ of the $W$- and $Z$-peak respectively (see
Table~\ref{tableIII}), but there are still
sufficient number of signal events to
affect the absolute normalization of the $W$ and $Z$ peaks.  When the Higgs
peak is isolated from the $W$ or $Z$ peak, the Higgs discovery, which only
depends on the actual number of signal events, should be viable up to about
$m_H=150$~GeV.
However, as $m_H$ increases from 150~GeV, the $B(H\rightarrow jj)$
drops very sharply, therefore the number of signal events becomes marginal for
discovery, and after 160~GeV there are too few signal events.

Here we can also estimate the effect of overall acceptance and detection
efficiencies of jets, say 25\% overall.  We should still have sufficient
number of signal events and large enough significance to cover the whole range
of 60 -- 150~GeV, except for $m_H$ right at $m_W$ where the significance goes
down below 6 to 3.8 (5.5) at $\sqrt{s}=1(2)$~TeV, and for $m_H=m_Z$ where the
significance goes down to 5.5 at $\sqrt{s}=1$~TeV.   The first exception
should be cleaned up because $m_H\simeq m_W$ will be covered with
ease at LEPII.   The second exception is only slightly below our requirement
of 6, so a slight increase in overall efficiency or $\sqrt{s}$  can solve.

\section{Conclusions and Discussions}
\label{secIV}

(i) We have done a signal-background analysis of the IMH search via the
channel $e^-\gamma\rightarrow W^-H\nu\rightarrow (jj) (b\bar b) \nu$ with and
without considering $b$-identification,
in a TeV $e^-\gamma$ collider, in which the
photon beam is realized by the  laser back-scattering method.  The continuum
backgrounds come from $e^-\gamma\rightarrow \bar tb\nu \rightarrow b\bar b\nu$
and $t\bar te^-\rightarrow b\bar b WWe^-$, while the discrete backgrounds come
{} from $e^-\gamma\rightarrow W^-Z\nu,\,WWe^-$ and $ZZe^-$.
We showed the results at both 1 and 2~TeV $e^+e^-$ machines, between which the
$WH\nu$ production is  large enough for IMH discovery.
However, at 0.5~TeV the $WH\nu$ production is too small for any realistic Higgs
search.

(ii) With 100\% $b$-identification the discovery of Higgs for the whole range
of $m_H=60-150$~GeV (marginally upto 160~GeV) should be viable at both
$\sqrt{s_{e^+e^-}}=1$ and 2~TeV.   With
$m_H\simeq m_Z$, since the signal is slightly larger than the background, the
exact absolute normalization of the $Z$-peak is not important.

(iii) Without $b$-identification the whole range of $m_H=60-150$~GeV
(marginally upto 160~GeV) should be covered at both energies.
With $m_H\simeq m_W$ or $m_Z$
the background is several times larger, therefore absolute normalization of
the $W$ and $Z$ peaks is important.  Fortunately, there are sufficient number
of signal events to affect the absolute normalization.

(iv) All the cross-sections in the Tables are assuming that the
jets are recognized with
100\% efficiency.   We also tried to estimate the effect of 25\% overall
acceptance and detection efficiencies.  In this case, the signal and
background events go down to 25\%, and the significance $S/\sqrt{B}$ is
halved.  As mentioned in previous section, even with this overall
efficiencies, the whole range of $m_H=60-150$~GeV should be covered for both
cases: with and without  considering $b$-identification.

(v) In the real collider experiment, the $b$-identification efficiency
will be somewhere  between our two extreme cases.  Therefore, we expect the
Higgs discovery should be viable for the whole range of $m_H=60-150$~GeV
between $\sqrt{s_{e^+e^-}}=1$ and 2~TeV inclusively,
provided that the absolute normalization of the $W$ and $Z$ peak is known to a
certain accuracy.  At $\sqrt{s_{e^+e^-}}=2$~TeV, the search for the IMH is
actually doing a little better, though not much, than that at 1~TeV because it
has about twice the signal, but also twice the discrete backgrounds, and
slightly less continuum background.

(v) In estimating the continuum background we take the invariant-mass $m(b\bar
b)$ or $m(jj)$ in the interval $m_H\pm 5$~GeV.  Due to limitations of the
detector we may not able to achieve this resolution, then we have to relax
this stringent requirement by some extent.  For example, if we take $m_H\pm
10$~GeV, which is quite conservative, the background coming from the continuum
increases by a factor of 2, because the continuum background is rather flat
(see Figs.~\ref{mbb-yes} and \ref{mbb-no}).  In this case, the significance of
the isolated Higgs-signal is reduced by $\sqrt{2}$.  From Tables~\ref{tableII}
and \ref{tableIII} we can see that even though the significance of the signal
(away from the $W$ or $Z$-peak) is reduced by
such factor, it is  still large enough for Higgs discovery.

(vi) In calculating the contribution from the resolved photon processes, we
used the DG parameterization \cite{DG} for the photon structure function.  DG
has  a relatively soft gluon spectrum.  If we choose LAC3 \cite{LAC3}
parameterization,
which has a relatively harder gluon spectrum, the contribution from resolved
photon process  is expected to increase by  a factor of  2 -- 3.  Even that the
continuum is still far below the Higgs-signal peak (see Figs.~\ref{mbb-yes} and
\ref{mbb-no}), so   this will not affect  our conclusions.

(vii) the effective $e\gamma$ luminosity might be less than the original
electron-positron luminosity \cite{teln}.  This fact will reduce our signal
and backgrounds by the same amount, but will  reduce the significance of the
signal, thus making the discovery of Higgs boson more difficult.  However,
this channel is still useful because in the future there are likely
improvements in the machine design that can optimize the effective luminosity.

\acknowledgements
This work was supported by the U.~S. Department of Energy, Division of
High Energy Physics, under Grant DE-FG02-91-ER40684.

\newpage
\appendix{}
\label{app}

In this appendix we present the matrix elements for
processes $e^-\gamma\rightarrow W^-H\nu,\,W^-Z\nu,\, W^-W^+e^-,\, ZZe^-,\,
\bar tb\nu$ and $t\bar te^-$, from
which explicit helicity amplitudes can be directly computed.
To start with, we introduce some general notation:
\begin{eqnarray}
g_a^W(f) & = & -g_b^W(f) = \frac{g}{2 \sqrt{2}} \, , \\
g_a^Z(f) & = & g_Z \left( {T_{3f}\over2} - Q_f x_{\rm w}\right) \, ,\\
g_b^Z(f) & = & - g_Z {T_{3f}\over2} \, ,\\
g_a^\gamma(f) & = & e Q_f\, ,\\
g_b^\gamma(f) & = & 0\, ,\\
g^V(f) & = & g_a^V(f) + g_b^V(f) \gamma^5\,\qquad(V=\gamma,W,Z)\,,\\
D^X(k) & = & \frac{1}{k^2-M_X^2 + i \Gamma_X(k^2) m_X}\,,\qquad
\Gamma_X(k^2) = \Gamma_X \theta(k^2) \nonumber \\
&&  \qquad (\mbox{with }X=\gamma,W,Z,H) \, ,\\
P_V^{\alpha \beta}(k) & = &  \left [ g^{\alpha \beta} + \frac{(1-\xi)k^\alpha
k^\beta}{\xi k^2 - m_V^2} \right ] D^V(k) \,, \\
\Gamma^\alpha (k_1,k_2;\epsilon_1,\epsilon_2) & = & (k_1-k_2)^\alpha \epsilon_1
\cdot \epsilon_2 + (2k_2+k_1) \cdot \epsilon_1 \epsilon_2^\alpha
- (2k_1+k_2) \cdot \epsilon_2 \epsilon_1^\alpha\, , \\
g_{VWW} & = & \left \{
               \begin{array}{ll}
                e \cot \theta_{\rm w}  \quad & {\rm for\ } V=Z \\
                e                      & {\rm for\ } V=\gamma \, .
               \end{array} \right.  \\ \nonumber
\end{eqnarray}
Here $Q_f$ and $T_{3f}$ are the electric charge (in units of the positron
charge) and the third component of weak isospin of the fermion $f$, $g$ is
the SU(2) gauge coupling, and $g_Z=g/\cos \theta_{\rm w}$,
$x_{\rm w}=\sin^2 \theta_{\rm w}$, with $\theta_{\rm w}$ being the weak
mixing angle in the Standard Model.  Dots between 4-vectors denote scalar
products and $g_{\alpha \beta}$ is the Minkowskian metric tensor with
$g_{00}=-g_{11}=-g_{22}=-g_{33}=1$; $\xi$ is a gauge-fixing parameter.

In Figs.~\ref{fey-wh} and \ref{fey-wz}, the momentum-labels $p_i$ denote
the momenta flowing along
the corresponding fermion lines in the direction of the arrows.
We shall denote the associated spinors by the symbols $u(p_i)$ and $\bar
u(p_i)$ for the incoming and outgoing arrows, which is usual for the
annihilation and creation of fermions, respectively.  In Fig.~\ref{fey-bt}
there is also creation of anti-fermion (corresponding to an incoming arrow
labeled by negative momentum $-p_i$), we shall denote its associated
spinor by $v(p_i)$.

\subsection{$e^-\gamma\rightarrow W^-H\nu$}

The contributing Feynman diagrams for $e^- (p_1) \gamma (p_2) \rightarrow
W^- (k_1) H(k_2) \nu (q_1)$  are shown in Fig.~\ref{fey-wh}.
We define a shorthand notation
\begin{equation}
J^\mu_1 = \bar u(q_1) \gamma^\mu g^W(e) u(p_1) \times D^W(p_1-q_1)\,,
\end{equation}
then the helicity amplitudes are given by
\begin{eqnarray}
{\cal M}^{(a)} &=& g^2 m_W \sin\theta_{\rm w}\,P^{\alpha\beta}_W(p_2-k_1) \,
\Gamma_\alpha\left( -k_1,\,p_2;\, \epsilon(k_1),\, \epsilon(p_2)\right )\,
   J_{1 \beta} \,,  \\
{\cal M}^{(b)} &=& -\,g^2 m_W \sin\theta_{\rm w}\, \epsilon(p_2) \cdot
\epsilon(k_1)\, k_2 \cdot J_1   \,
\frac{\xi}{\xi(p_2-k_1)^2 - m_W^2} \,,  \\
{\cal M}^{(c)} &=& g^2 m_W \sin\theta_{\rm w}\,
P_W^{\alpha\beta}(k_1+k_2) \,
\Gamma_\alpha(p_2,\,p_1-q_1;\, \epsilon(p_2),\, J_1)\,
         \epsilon_\beta(k_1) \,,\\
{\cal M}^{(d)} &=& g^2 m_W \sin\theta_{\rm w}\, \epsilon(p_2) \cdot J_1 \;
k_2 \cdot \epsilon(k_1)
\, \frac{\xi}{\xi(k_1+k_2)^2 - m_W^2} \,, \\
{\cal M}^{(e)} &=& -g m_W \,P^{\alpha\beta}_W(k_1+k_2) \epsilon_\alpha(k_1)
\nonumber \\
&& \quad \times \; \bar u(q_1) \gamma_\beta g^W(e) \frac{ \overlay{/}{p}_1 +
\overlay{/}{p}_2 + m_e} {(p_1+p_2)^2 - m_e^2} \overlay{/}{\epsilon}(p_2)
g^\gamma(e) u(p_1)  \,,
\end{eqnarray}

\subsection{$e^-\gamma\rightarrow W^-Z \nu$}

The contributing Feynman diagrams for $e^- (p_1) \gamma (p_2)
\rightarrow W^-(k_1) Z(k_2) \nu (q_1)$ are given in Fig.~\ref{fey-wz}(a).
We define a shorthand notation
\begin{equation}
\begin{array}{rcl}
J^\mu_1 &=& \bar u(q_1) \gamma^\mu g^W(e) u(p_1) \times D^W(p_1-q_1)\,,
\end{array}
\end{equation}
then the  helicity amplitudes are given by
\begin{eqnarray}
{\cal M}^{(a)} &=& -\, g_{ZWW} g_{\gamma WW} \,
\Gamma_\alpha(-k_1,\,p_2;\, \epsilon(k_1), \, \epsilon(p_2) ) \,
P^{\alpha\beta}_W(p_2-k_1) \nonumber \\
&& \qquad \times \, \Gamma_\beta(-k_2,\,p_1-q_1;\, \epsilon(k_2), \, J_1 )\,,\\
{\cal M}^{(b)} &=& -\, g_{ZWW} g_{\gamma WW} \,
\Gamma_\alpha( k_2,\,k_1;\, \epsilon(k_2), \, \epsilon(k_1) ) \,
P^{\alpha\beta}_W(k_1+k_2) \nonumber  \\
&& \qquad \times \,\Gamma_\beta( p_2,\,p_1-q_1;\, \epsilon(p_2), \, J_1 ) \,,\\
{\cal M}^{(c)} &=& g_{ZWW} g_{\gamma WW} \left[
  2 \epsilon(p_2) \cdot \epsilon(k_2) \epsilon(k_1)\cdot J_1
- \epsilon(p_2) \cdot J_1  \epsilon(k_1)\cdot \epsilon(k_2)\right. \nonumber \\
&& \left. \qquad \qquad \qquad - \,  \epsilon(p_2) \cdot \epsilon(k_1)
              \epsilon(k_2)\cdot J_1          \right] \,, \\
{\cal M}^{(d,e)} &=& g_{\gamma WW} \Gamma_\alpha(-k_1,\,p_2;\,\epsilon(k_1),\,
\epsilon(p_2) ) P^{\alpha\beta}_W(p_2- k_1)  \nonumber \\
&& \times \left [ \bar u(q_1) \gamma_\beta g^W (e) \frac{\overlay{/}{p}_1 -
\overlay{/}{k}_2 + m_e }{(p_1-k_2)^2 - m^2_e}\, \overlay{/}{\epsilon}(k_2)
g^Z(e) u(p_1)  \right. \nonumber \\
&& \qquad \qquad \left. + \; \bar u(q_1) \overlay{/}{\epsilon}(k_2) g^Z(\nu)
\frac{\overlay{/}{q}_1 + \overlay{/}{k}_2}{(q_1+k_2)^2}\, \gamma_\beta g^W(e)
u(p_1) \right ]\,, \\
{\cal M}^{(f)} &=& g_{ZWW} \Gamma_\alpha( k_2,\,k_1;\,\epsilon(k_2),\,
\epsilon(k_1)) P^{\alpha\beta}_W(k_1+k_2) \nonumber \\
&& \times\; \bar u(q_1) \gamma_\beta g^W(e) \frac{\overlay{/}{p}_1 +
\overlay{/}{p}_2 + m_e}{(p_1+p_2)^2 - m_e^2}\, \overlay{/}{\epsilon}(p_2)
g^\gamma(e) u(p_1)\,, \\
{\cal M}^{(g)} &=& - \bar u(q_1) \overlay{/}{\epsilon}(k_1) g^W(e)
\frac{ \overlay{/}{q}_1 + \overlay{/}{k}_1 + m_e}{(q_1+k_1)^2 - m_e^2}
\overlay{/}{\epsilon}(k_2) g^Z(e)
\frac{ \overlay{/}{p}_1 + \overlay{/}{p}_2 + m_e}{(p_1+p_2)^2 -m_e^2}
\nonumber \\
&& \qquad \overlay{/}{\epsilon}(p_2) g^\gamma(e) u(p_1) \,, \\
{\cal M}^{(h)} &=& - \bar u(q_1) \overlay{/}{\epsilon}(k_1) g^W(e)
\frac{ \overlay{/}{q}_1 + \overlay{/}{k}_1 + m_e}{(q_1+k_1)^2 - m_e^2}
\overlay{/}{\epsilon}(p_2) g^\gamma(e)
\frac{ \overlay{/}{p}_1 - \overlay{/}{k}_2 + m_e}{(p_1-k_2)^2 -m_e^2}
\nonumber \\
&& \qquad \overlay{/}{\epsilon}(k_2) g^Z(e) u(p_1) \,, \\
{\cal M}^{(i)} &=& - \bar u(q_1) \overlay{/}{\epsilon}(k_2) g^Z(\nu)
\frac{ \overlay{/}{q}_1 + \overlay{/}{k}_2}{(q_1+k_2)^2}
\overlay{/}{\epsilon}(k_1) g^W(e)
\frac{ \overlay{/}{p}_1 + \overlay{/}{p}_2 + m_e}{(p_1+p_2)^2 -m_e^2}
\nonumber \\
&& \qquad \overlay{/}{\epsilon}(p_2) g^\gamma(e) u(p_1) \,, \\
{\cal M}^{(j)} &=& - g^2 m_W^2 x_{\rm w} \tan\theta_{\rm w}
\frac{\xi}{\xi(p_2-k_1)^2 -m_W^2}\, \epsilon(k_1) \cdot \epsilon(p_2) \,
               \epsilon(k_2) \cdot J_1 \,\\
{\cal M}^{(k)} &=& - g^2 m_W^2 x_{\rm w} \tan\theta_{\rm w}
\frac{\xi}{\xi(k_1+k_2)^2 -m_W^2} \, \epsilon(k_1) \cdot \epsilon(k_2) \,
               \epsilon(p_2) \cdot J_1 \,.
\end{eqnarray}

\subsection{$e^-\gamma \rightarrow W^- W^+ e^-$}

The contributing Feynman diagrams for the process
$e^-(p_1)\gamma(p_2) \rightarrow W^-(k_1) W^+(k_2) e^-(q_1)$ are shown
in Fig.~\ref{fey-wz}(b).  We can also  define a shorthand notation
\begin{equation}
\begin{array}{rcl}
J^\mu_V &=& \bar u(q_1) \gamma^\mu g^V(e) u(p_1) \times D^V(p_1-q_1)\,,
\quad {\rm where}\; V=\gamma,\,Z
\end{array}
\end{equation}
then the  helicity amplitudes are given by
\begin{eqnarray}
{\cal M}^{(a)} &=& \sum_{V=\gamma,Z} -\,g_{VWW}\, g_{\gamma WW} \,
P^{\alpha\beta}_W(p_2-k_2) \nonumber \\
&& \quad \times \,\Gamma_\alpha(-k_1,\, p_1-q_1;\, \epsilon(k_1),\, J_V)
       \, \Gamma_\beta( p_2,\, -k_2;\, \epsilon(p_2),\, \epsilon(k_2) ) \,, \\
%
%
{\cal M}^{(b)} &=& \sum_{V=\gamma,Z} - \,g_{VWW} \, g_{\gamma WW} \,
P^{\alpha\beta}_W(p_2-k_1) \nonumber \\
&& \quad \times \, \Gamma_\alpha(p_1-q_1,\, -k_2;\, J_V,\, \epsilon(k_2) )
       \, \Gamma_\beta( -k_1,\, p_2;\, \epsilon(k_1),\, \epsilon(p_2) ) \,, \\
{\cal M}^{(c)} &=& \sum_{V=\gamma,Z} g_{VWW} g_{\gamma WW} \,\left [
2 \epsilon(k_1) \cdot \epsilon(k_2)\, \epsilon(p_2) \cdot J_V \nonumber
\right. \nonumber \\
&& \left. \qquad\qquad -\,\epsilon(k_1) \cdot J_V \, \epsilon(k_2)\cdot
          \epsilon(p_2)
          -\, \epsilon(k_2) \cdot J_V \, \epsilon(k_1)\cdot \epsilon(p_2)\,
    \right ] \,, \\
{\cal M}^{(d)} &=& - \bar u(q_1) \overlay{/}{\epsilon}(k_2) g^W(e)
\frac{\overlay{/}{q}_1 + \overlay{/}{k}_2}{(q_1+k_2)^2}
\overlay{/}{\epsilon}(k_1) g^W(e)
\frac{\overlay{/}{p}_1 + \overlay{/}{p}_2 + m_e}{(p_1+p_2)^2 - m_e^2}
\nonumber \\
&& \qquad \overlay{/}{\epsilon}(p_2) g^\gamma(e) u(p_1) \,, \\
{\cal M}^{(e)} &=& - \bar u(q_1) \overlay{/}{\epsilon}(p_2) g^\gamma(e)
\frac{\overlay{/}{q}_1 - \overlay{/}{p}_2 + m_e}{(q_1-p_2)^2 - m_e^2}
\overlay{/}{\epsilon}(k_2) g^W(e)
\frac{\overlay{/}{p}_1 - \overlay{/}{k}_1 }{(p_1 - k_1)^2} \nonumber \\
&& \qquad \overlay{/}{\epsilon}(k_1) g^W(e) u(p_1) \,, \\
{\cal M}^{(f)} &=& \sum_{V=\gamma,Z} g_{VWW} D^V(k_1+k_2) \,
\Gamma_\alpha (k_1,\,k_2;\, \epsilon(k_1),\, \epsilon(k_2) ) \nonumber \\
&& \quad  \times \,\bar u(q_1) \gamma^\alpha g^V(e)
\frac{\overlay{/}{p}_1 + \overlay{/}{p}_2 + m_e }{(p_1 + p_2)^2 - m_e^2} \,
\overlay{/}{\epsilon}(p_2) g^\gamma(e) u(p_1) \,, \\
{\cal M}^{(g)} &=& \sum_{V=\gamma,Z} g_{VWW} D^V(k_1+k_2) \,
\Gamma_\alpha (k_1,\,k_2;\, \epsilon(k_1),\, \epsilon(k_2) ) \nonumber \\
&& \quad \times \, \bar u(q_1) \overlay{/}{\epsilon}(p_2) g^\gamma(e)
\frac{\overlay{/}{q}_1 - \overlay{/}{p}_2 + m_e }{(q_1 - p_2)^2 - m_e^2} \,
\gamma^\alpha  g^V(e)  u(p_1) \,, \\
{\cal M}^{(h)} &=& g_{\gamma WW} P^{\alpha\beta}_W(p_2-k_2)
\Gamma_\alpha ( p_2,\, -k_2;\, \epsilon(p_2),\, \epsilon(k_2) ) \nonumber \\
&& \quad \times \, \bar u(q_1) \gamma_\beta g^W(e)
\frac{\overlay{/}{p}_1 - \overlay{/}{k}_1}{(p_1 - k_1 )^2} \,
\overlay{/}{\epsilon}(k_1) g^W(e) u(p_1) \,, \\
{\cal M}^{(i)} &=& g_{\gamma WW} P^{\alpha\beta}_W(p_2-k_1) \,
\Gamma_\alpha ( -k_1,\, p_2;\, \epsilon(k_1),\, \epsilon(p_2) ) \nonumber \\
&& \quad \times \, \bar u(q_1)   \overlay{/}{\epsilon}(k_2) g^W(e)
\frac{\overlay{/}{q}_1 + \overlay{/}{k}_2}{(q_1 + k_2 )^2} \,
\gamma_\beta  g^W(e) u(p_1) \,, \\
{\cal M}^{(j)} &=& \sum_{V=\gamma,Z} g^2 m_W^2 x_{\rm w} \,
\frac{\xi}{\xi(p_2-k_2)^2 - m_W^2}\, \epsilon(p_2) \cdot \epsilon(k_2) \,
\epsilon(k_1) \cdot J_V \nonumber \\
&& \qquad \times \, \left \{ \begin{array}{ll}
                              -\tan\theta_{\rm w} & \quad {\rm for}\; V=Z \\
                              1      & \quad {\rm for}\; V=\gamma
                            \end{array}
                   \right.
    \,, \\
{\cal M}^{(k)} &=& \sum_{V=\gamma,Z} g^2 m_W^2 x_{\rm w} \,
\frac{\xi}{\xi(p_2-k_1)^2 - m_W^2}\, \epsilon(p_2) \cdot \epsilon(k_1) \,
\epsilon(k_2) \cdot J_V \nonumber \\
&& \qquad \times \, \left \{ \begin{array}{ll}
                              -\tan\theta_{\rm w} & \quad {\rm for}\; V=Z \\
                              1      & \quad {\rm for}\; V=\gamma
                            \end{array}
                   \right.
    \,.
\end{eqnarray}

\subsection{$e^-\gamma\rightarrow ZZ e^-$}

The contributing Feynman diagrams for the process $e^-(p_1)\gamma(p_2)
 \rightarrow Z(k_1) Z(k_2) e^-(q_1)$ are the same as the diagram (e) in
Fig.~\ref{fey-wz}(b) with the $W$-bosons replaced by $Z$-bosons
plus  all possible permutations.  Totally it has six
contributing Feynman diagrams.  They are given by
\begin{eqnarray}
{\cal M}^{(a)} &=& - \bar u(q_1) \overlay{/}{\epsilon}(k_1) g^Z(e)
\frac{\overlay{/}{q}_1 + \overlay{/}{k}_1 +m_e }{(q_1+k_1)^2 - m_e^2 }
\overlay{/}{\epsilon}(k_2) g^Z(e)
\frac{\overlay{/}{p}_1 + \overlay{/}{p}_2 +m_e }{(p_1 + p_2)^2 - m_e^2 }
\nonumber \\
&& \qquad \overlay{/}{\epsilon}(p_2) g^\gamma(e) u(p_1) \,, \\
{\cal M}^{(b)} &=& - \bar u(q_1) \overlay{/}{\epsilon}(k_1) g^Z(e)
\frac{\overlay{/}{q}_1 + \overlay{/}{k}_1 +m_e }{(q_1+k_1)^2 - m_e^2 }
\overlay{/}{\epsilon}(p_2) g^\gamma(e)
\frac{\overlay{/}{p}_1 - \overlay{/}{k}_2 +m_e }{(p_1 - k_2)^2 - m_e^2 }
\nonumber \\
&& \qquad \overlay{/}{\epsilon}(k_2) g^Z(e) u(p_1) \,, \\
{\cal M}^{(c)} &=& - \bar u(q_1) \overlay{/}{\epsilon}(p_2) g^\gamma(e)
\frac{\overlay{/}{q}_1 - \overlay{/}{p}_2 +m_e }{(q_1-p_2)^2 - m_e^2 }
\overlay{/}{\epsilon}(k_1) g^Z(e)
\frac{\overlay{/}{p}_1 - \overlay{/}{k}_2 +m_e }{(p_1 - k_2)^2 - m_e^2 }
\nonumber \\
&& \qquad \overlay{/}{\epsilon}(k_2) g^Z(e) u(p_1) \,,
\end{eqnarray}
plus those terms with $(k_1 \leftrightarrow k_2)$.

\subsection{$e^-\gamma\rightarrow \bar t b\nu$}

The contributing Feynman diagrams for $e^-(p_1) \gamma (p_2) \rightarrow
\bar t(k_1) b(k_2) \nu (q_1)$ are shown in Fig.~\ref{fey-bt}(a).  We
define the following shorthand notation:
\begin{equation}
J^\mu_1  = \bar u(q_1)\gamma^\mu g^W(e) \,u(p_1) \times D^W(p_1-q_1)\,,\\
\end{equation}
then the helicity amplitudes are given by
\begin{eqnarray}
{\cal M}^{(a)} &=& \bar u(k_2)\, \overlay{/}{\epsilon}(p_2)\, g^\gamma(b)\,
   \frac{ \overlay{/}{k}_2 - \overlay{/}{p}_2 + m_b} {(k_2-p_2)^2 - m_b^2}\,
   \overlay{/}{J}_1 \, g^W(t)\, v(k_1) \,, \\
{\cal M}^{(b)} &=& \bar u(k_2)\, \overlay{/}{J}_1 \, g^W(t)\,
   \frac{\overlay{/}{p}_2 -\overlay{/}{k}_1 + m_t} {(p_2-k_1)^2 - m_t^2}\,
   \overlay{/}{\epsilon}(p_2)\, g^\gamma(t)\, v(k_1) \,, \\
{\cal M}^{(c)} &=& - g \sin\theta_{\rm w}\, P^{\alpha\beta}_W(k_1 +k_2) \,
 \bar u(k_2) \gamma_\alpha g^W(t) v(k_1) \nonumber \\
&& \quad \times  \Gamma_\beta(p_2,\,p_1-q_1;\, \epsilon(p_2),\, J_1 )\,, \\
{\cal M}^{(d)} &=& \frac{g^2}{2\sqrt{2}}\, \sin\theta_{\rm w} \,
   \epsilon(p_2)\cdot J_1 \, \frac{\xi}{\xi(k_1+k_2)^2 - m_W^2} \,
\nonumber \\
&& \times \;
      \bar u(k_2) \left[(m_t-m_b) + (m_t+m_b)\gamma^5 \right] v(k_1) \,, \\
{\cal M}^{(e)} &=& \bar u(q_1) \gamma_\alpha g^W(e)
  \frac{\overlay{/}{p}_1 + \overlay{/}{p}_2 + m_e}{(p_1+p_2)^2 -m_e^2} \,
  \overlay{/}{\epsilon}(p_2) g^\gamma(e) \, u(p_1) \,\nonumber \\
&& \times \; P^{\alpha\beta}_W(k_1+k_2) \, \bar u(k_2) \gamma_\beta g^W(t) \,
v(k_1)\,.
\end{eqnarray}
The diagrams for the ``resolved" photon process $e^-g\rightarrow \bar t b\nu$
are the same  as diagrams (a) and (b) in Fig.~\ref{fey-bt}(a), with the photon
replaced by the gluon.  The helicity amplitudes are the same except with the
coupling $g^\gamma$ replaced by $g_s$, and a different color factor.  The
color factor to multiply the matrix element squared is 3 and 1/2 for the
$e\gamma$ and the resolved process, respectively.

\subsection{$e^-\gamma\rightarrow t\bar te^-$}

The contributing Feynman diagrams for $e^-(p_1) \gamma(p_2) \rightarrow
\bar t(k_1) t(k_2) e^-(q_1)$  are shown in Fig.~\ref{fey-bt}(b).  We also
define the shorthand notations:
\begin{equation}
\begin{array}{rcl}
J^\mu_{Ve} &=& \bar u(q_1) \gamma^\mu g^V(e) u(p_1) \times D^V(p_1-q_1)\,,\\
J^\mu_{Vt} &=& \bar u(k_2) \gamma^\mu g^V(t) v(k_1) \times D^V(k_1+k_2)\,,
\end{array}
\end{equation}
then the helicity amplitudes are given by
\begin{eqnarray}
{\cal M}^{(a)} &=& \sum_{V=\gamma,Z}
    \bar u(k_2) \overlay{/}{\epsilon}(p_2) g^\gamma(t)
  \frac{\overlay{/}{k}_2 - \overlay{/}{p}_2 + m_t} {(k_2-p_2)^2 -m_t^2}
  \overlay{/}{J}_{Ve} g^V(t) v(k_1) \,, \\
{\cal M}^{(b)} &=& \sum_{V=\gamma,Z} \bar u(k_2)\, \overlay{/}{J}_{Ve} g^V(t)
\,   \frac{\overlay{/}{p}_2 -\overlay{/}{k}_1 + m_t} {(p_2-k_1)^2 - m_t^2}\,
   \overlay{/}{\epsilon}(p_2) g^\gamma(t)\, v(k_1) \,, \\
{\cal M}^{(c)} &=& \sum_{V=\gamma,Z} \bar u(q_1) \,\overlay{/}{J}_{Vt} g^V(e)
\,  \frac{\overlay{/}{p}_1 + \overlay{/}{p}_2 + m_e}{(p_1+p_2)^2 -m_e^2} \,
  \overlay{/}{\epsilon}(p_2) g^\gamma(e)\, u(p_1) \,,\\
{\cal M}^{(d)} &=& \sum_{V=\gamma,Z}
  \bar u(q_1)\, \overlay{/}{\epsilon}(p_2) g^\gamma(e) \,
  \frac{\overlay{/}{q}_1 - \overlay{/}{p}_2 + m_e}{(q_1-p_2)^2 -m_e^2} \,
  \overlay{/}{J}_{Vt} g^V(e) \,u(p_1) \,.
\end{eqnarray}
The color factor to multiply the matrix element squared is 3 for this process.

These matrix elements are to be squared, summed over polarizations
and spins of the final state gauge-bosons and fermions respectively,
and then averaged  over the
polarizations of the incoming photon and spins of the initial state electron.
Then the cross section $\sigma$ is obtained by folding the subprocess
cross-section $\hat \sigma$ in with the photon luminosity function as
\begin{equation}
\sigma(s) = \int^{x_{max}}_{M_{\rm final}/s} dx \;F_{\gamma/e}(x)
\; \hat \sigma( \hat s=xs)  \,,
\end{equation}
where
\begin{equation}
\begin{array}{rcl}
\hat\sigma(\hat s) &=&  \frac{1}{2 (\hat s -m_e^2)}   \int
                     \frac{d^3 k_1}{(2\pi)^3 k_1^0} \,
                     \frac{d^3 k_2}{(2\pi)^3 k_2^0}\,
                     \frac{d^3 q_1}{(2\pi)^3 q_1^0}  \\
 && \qquad \quad \times \;    (2\pi)^4 \delta^{(4)} (p_1+p_2-k_1-k_2-q_1) \,
               \sum | {\cal M}|^2 \, \\
\end{array}
\end{equation}
and $M_{\rm final}$ is the sum of the masses of the final state particles.


\begin{table}
\caption{\label{tableI}Table showing the effectiveness of various combinations
of the cuts in Eqs.~(\ref{ptvvcut}), (\ref{vetoing}) and (\ref{cosvvcut}) for
$\sqrt{s_{e^+e^-}}=1(2)$~TeV, with $m_t=150$ and $m_H=100$~GeV.
The cross-sections are given in the unit of fb.  No branching fractions of the
bosons are included.  The numbers in the parenthesis are for
$\sqrt{s_{e^+e^-}} = 2$~TeV. }
\begin{tabular}{||l|@{\extracolsep{-0.2in}}ccccccc||}
Combinations & $WH$  & $WZ$ & $WW$ & $ZZ$  &$e^-\gamma\rightarrow b\bar t\nu$&
$e^-g\rightarrow b\bar t\nu$  &  $t\bar te^-$ \\
\hline
(a) No cuts  & 79 &  390   & 10400   & 10    & 55 & 8.1 & 83    \\
            & (200) & (970) & (19500) & (6.8) & (91)& (43) & (160) \\
&&&&&&& \\
(b) Eq.~(\ref{ptvvcut}) & 73 & 380 & 1960 & 2.7 & 52 & 7.2 &  79\\
                      & (160) & (930) & (2960) & (1.9) & (79)& (31) & (150) \\
&&&&&&& \\
(c) Eqs.~(\ref{ptvvcut}) and (\ref{vetoing})
       & 73 & 380 & 900 & 0.54  & 52 & 7.2 &  66\\
       & (160) & (930) & (1800) & (0.24) & (79)& (31) &  (130) \\
&&&&&&& \\
(d) Eqs.~(\ref{ptvvcut}), (\ref{vetoing})
      & 55 & 250 & 510 & 0.19 & 20 & 2.4 &  22  \\
 \mbox{\hspace{0.3in}}and (\ref{cosvvcut})
& (110) & (560) & (1000) & (0.083) & (22)& (7.7) &  (22) \\
&&&&&&& \\
\end{tabular}
\end{table}
\begin{table}
\caption{\label{tableII}Cross-sections (fb) for the signal
and various backgrounds, total background $B$, and the significance
$S/\sqrt{B}$ (integrated luminosity=10 fb$^{-1}$) of the
signal for $m_H=60-160$~GeV at $\sqrt{s_{e^+e^-}}=1(2)$~TeV.  Here the
acceptance cuts are Eqs.~(\ref{ptvvcut}), (\ref{vetoing}) and (\ref{cosvvcut}).
The discrete backgrounds are calculated using Eq.~(\ref{masscut})
and the continuum backgrounds  are with  $m(b\bar b)$ in between $m_H\pm
5$~GeV.
Also we assume 100\% $b$-tagging, $m_t=150$~GeV, and the branching
fractions of the $V_1 \rightarrow jj$ and $V_2 \rightarrow b\bar b$ are
included.
}
\renewcommand{\arraystretch}{0.8}
\begin{tabular}{||c|c|ccc|ccc|c|c||}
$m_H$ & signal & \multicolumn{3}{c|}{discrete backgrounds}  &
\multicolumn{3}{c|}{contiuum backgrounds} & total &$S/\sqrt{B}$
 \\
& $WH$ & $WZ$ & $WW$ & $ZZ$ & $\bar tb\nu$ & $e^-g\rightarrow \bar tb\nu$
 & $t\bar te^-$ & $B$&10 fb$^{-1}$ \\
\hline
60  & 40  & 0 & 0 & 0 & 0.56 & 0.18    & 0.89  & 1.6  & 99 \\
    & (75)&(0)&(0)&(0)&(0.34)& (0.40) & (0.67) & (1.4)& (200)\\
\hline
70  & 36  & 0 & 0 & 0 & 0.59 & 0.12 & 0.90     & 1.6  & 90 \\
    &(72) &(0)&(0)&(0)&(0.39)& (0.37)   &(0.74)& (1.5)& (186)\\
\hline
80  & 35  & 0 & 0 & 0 & 0.58 & 0.10 & 0.88 &  1.6   & 89\\
    & (69)&(0)&(0)&(0)&(0.41)& (0.32)  &(0.72)&(1.5)& (181)\\
\hline
90  & 33  & 23   & 0 & 0.035 & 0.56 & 0.083 & 0.83     &  25 & 21 \\
    &(67) & (52) &(0)&(0.015)&(0.41)& (0.28)    &(0.70) & (53)&(29) \\
\hline
100 & 31   & 3.0  & 0 & 0.0047 & 0.52 & 0.074 & 0.78      & 4.4 & 47 \\
    & (64) & (6.9)&(0)&(0.0020)&(0.41)& (0.24)    &(0.66) &(8.2)& (71)\\
\hline
110 & 28  & 0 & 0 & 0 & 0.52  & 0.055 & 0.74       & 1.3  & 77 \\
    &(59) &(0)&(0)&(0)&(0.41) & (0.21)    & (0.62) & (1.2)&(168) \\
\hline
120 & 24   & 0 & 0 & 0 & 0.50 & 0.045 & 0.70  &   1.2 &  68\\
    &(50.5)&(0)&(0)&(0)&(0.41)& (0.18)    &(0.58) &(1.2) &(148) \\
\hline
130 & 17.5 & 0 & 0 & 0 & 0.48 & 0.040 & 0.64       &1.2  & 51 \\
    & (38) &(0)&(0)&(0)&(0.41)& (0.15)    & (0.53) &(1.1)& (115)\\
\hline
140 & 11    & 0 & 0 & 0 & 0.47 & 0.034  & 0.61       & 1.1 & 33 \\
    &(24.5) &(0)&(0)&(0)&(0.41)&  (0.13)    & (0.51) &(1.1)& (76)\\
\hline
150 & 5.3  & 0 & 0 & 0 & 0.46 & 0.028  & 0.55     & 1.0  & 16 \\
    & (12) &(0)&(0)&(0)&(0.40)& (0.12)     &(0.47)&(0.99)& (38)\\
\hline
160 & 1.0  & 0 & 0 & 0 & 0.45  & 0.021  & 0.51       & 0.98 & 3.2\\
    &(2.4) &(0)&(0)&(0)&(0.39) & (0.10)     & (0.46) &(0.96)& (7.8)
\end{tabular}
\end{table}
%
\begin{table}
\caption{\label{tableIII}
Cross-sections (fb) for the signal
and various backgrounds, total background $B$, and the significance
$S/\sqrt{B}$ (integrated luminosity=10 fb$^{-1}$) of the
signal for $m_H=60-160$~GeV at $\sqrt{s_{e^+e^-}}=1(2)$~TeV.  Here the
acceptance cuts are Eqs.~(\ref{ptvvcut}), (\ref{vetoing}) and (\ref{cosvvcut}).
The discrete backgrounds are calculated using Eq.~(\ref{masscut})
and the continuum backgrounds are with $m(b\bar b)$ in between $m_H\pm 5$~GeV.
Also we assume NO $b$-identification, $m_t=150$~GeV, and the branching
fractions of the $V_1,\,V_2\rightarrow jj$ are included.
}
\renewcommand{\arraystretch}{0.8}
\begin{tabular}{||c|c|ccc|ccc|c|c||}
$m_H$ & signal & \multicolumn{3}{c|}{discrete backgrounds}  &
\multicolumn{3}{c|}{contiuum backgrounds} & total & $S/\sqrt{B}$\\
& $WH$ & $WZ$ & $WW$ & $ZZ$ & $\bar tb\nu$ & $e^-g\rightarrow \bar tb\nu$ &
  $t\bar te^-$ &$B$& 10fb$^{-1}$ \\
\hline
60  & 42  & 0 & 0 & 0 & 0.56 & 0.18 & 0.89         & 1.6 & 100 \\
    & (79.5)&(0)&(0)&(0)&(0.34)& (0.40)   & (0.67) &(1.4)& (210)\\
\hline
70  & 40  & 0 & 0 & 0 & 0.59 & 0.12 & 0.90     & 1.6 &  100\\
    &(77) &(0)&(0)&(0)&(0.39)& (0.37)   &(0.74)&(1.5)& (200)\\
\hline
80  & 38  & 0 & 250 & 0 & 0.58 & 0.10 & 0.88     & 250 & 7.6 \\
    & (74)&(0)&(490)&(0)&(0.41)& (0.32)   &(0.72)&(490)& (11)\\
\hline
90  & 36  & 110   & 0.14 & 0.082 & 0.56 & 0.083 & 0.83      & 110 & 11 \\
    &(71) & (240) &(0.28)&(0.036)&(0.41)& (0.28)    &(0.70) &(240)&(14) \\
\hline
100 & 33   & 14  & 0 & 0.011 & 0.52 & 0.074 & 0.78       & 15 & 27 \\
    & (68) & (32)&(0)&(0.0048)&(0.41)& (0.24)    &(0.66) & (33)&(37) \\
\hline
110 & 30  & 0 & 0 & 0 & 0.52  & 0.055 & 0.74       & 1.3 & 83\\
    &(63) &(0)&(0)&(0)&(0.41) & (0.21)    & (0.62) &(1.2)&(180) \\
\hline
120 & 25   & 0 & 0 & 0 & 0.50 & 0.045 & 0.70      & 1.2 & 71\\
    &(54)  &(0)&(0)&(0)&(0.41)& (0.18)    &(0.58) &(1.2)&(160) \\
\hline
130 & 19 & 0 & 0 & 0 & 0.48 & 0.040 & 0.64         & 1.2  &  56\\
    & (41) &(0)&(0)&(0)&(0.41)& (0.15)    & (0.53) &(1.1) &(120) \\
\hline
140 & 12    & 0 & 0 & 0 & 0.47 & 0.034  & 0.61     & 1.1 & 36 \\
    &(26) &(0)&(0)&(0)&(0.41)&  (0.13)    & (0.51) &(1.1)& (80) \\
\hline
150 & 5.6  & 0 & 0 & 0 & 0.46 & 0.028  & 0.55     & 1.0  & 17 \\
    & (13) &(0)&(0)&(0)&(0.40)& (0.12)     &(0.47)&(0.99)&(41) \\
\hline
160 & 1.1  & 0 & 0 & 0 & 0.45  & 0.021  & 0.51       & 0.98 & 3.5 \\
    &(2.6) &(0)&(0)&(0)&(0.39) & (0.10)     & (0.46) &(0.96)& (8.4)
\end{tabular}
\end{table}
%

\figure{\label{fey-wh}
Contributing Feynman diagrams for the process $e^-\gamma\rightarrow W^-H\nu$}

\figure{\label{fey-wz}
Contributing Feynman diagrams for the processes (a) $e^-\gamma\rightarrow
W^-Z\nu$, (b) $W^-W^+e^-$.}

\figure{\label{fey-bt}
Contributing Feynman diagrams for the processes (a) $e^-\gamma\rightarrow
\bar tb\nu$, (b) $t\bar te^-$.}

\figure{\label{cross}
Total cross sections in pb for signal and various backgrounds versus the
center-of-mass energies $\sqrt{s_{e^+e^-}}$
of the parent $e^+e^-$ collider for $m_H=100$~GeV and
$m_t=150$~GeV  before imposing any acceptance cuts.
No branching fractions are included.  The dash-dotted line represents the
resolved photon process for $e^-g(\gamma) \rightarrow \bar t b\nu$.}

\figure{\label{ptvv}
The dependence of the differential cross section $d\sigma/dp_T(VV)$ on the
transverse momentum of the boson-pair for signal and various backgrounds
at (a) $\sqrt{s_{e^+e^-}}=1~$TeV and (b)
2~TeV.  NO branching fractions of the boson-pair  are included here.}

\figure{\label{cosv1v2}
The dependence of the differential cross section $d\sigma/d \cos\theta_V$ on
the cosine of angle between the positive $z$-axis and the direction of (a)
boson $V_1$ and  (b) boson $V_2$, and (c) the differential cross-section
of $d\sigma/d \cos\theta_{V_1} \cos\theta_{V_2}$ versus
$\cos\theta_{V_1} \cos\theta_{V_2}$, at
$\sqrt{s_{e^+e^-}}=1$~TeV.  The acceptance cuts are $p_T(VV)>15$~GeV and
central $e^-$ vetoing.  NO branching fractions of the boson-pair are
included.}

\figure{\label{mbb-yes}
The dependence of the differential cross section $d\sigma/dm(b\bar b)$ on the
invariant mass of the $b\bar b$ pair (coming from $V_2$)
 for the signal and various backgrounds,
with the acceptance cuts of $p_T(VV)>15(30)$~GeV, central electron vetoing
and $\cos\theta_{V_1} \cos\theta_{V_2} <0$,
at (a) $\sqrt{s_{e^+e^-}}=1$~TeV, and (b) 2~TeV.  The branching fractions of
the $W,Z \rightarrow jj$ and $H,Z\rightarrow b\bar b$ are included.
We assumed a 100\% $b$-identification.}

\figure{\label{mbb-no}
The dependence of the differential cross section $d\sigma/dm(jj)$ on the
invariant mass of the $jj$ (coming from $V_2$) pair for the signal
and various backgrounds, with the acceptance cuts of
$p_T(VV)>15(30)$~GeV, central electron vetoing
and $\cos\theta_{V_1} \cos\theta_{V_2} <0$,
at (a) $\sqrt{s_{e^+e^-}}=1$~TeV, and (b) 2~TeV.  The branching fractions of
the $W,Z,H \rightarrow jj$ are included.  Here we did NOT assume
$b$-tagging.}

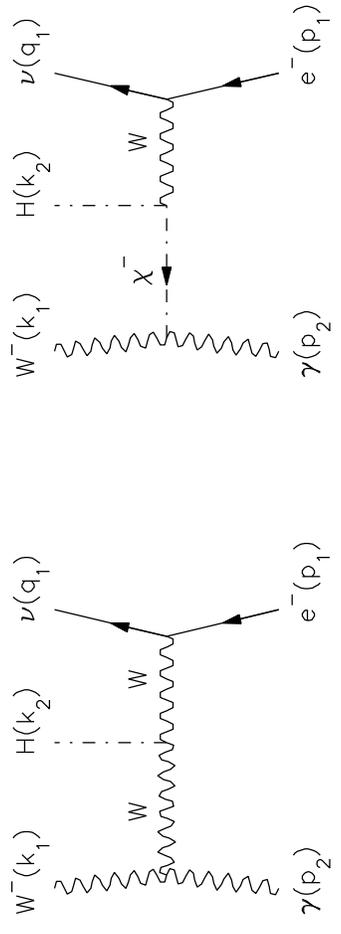
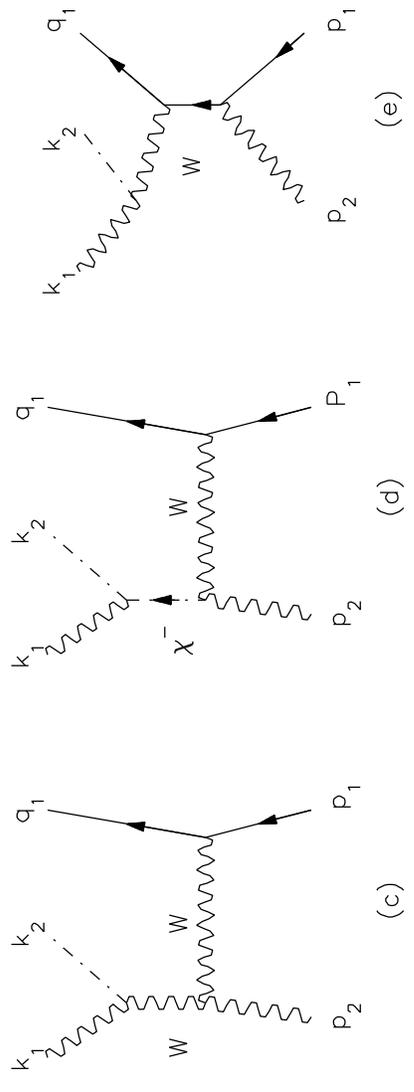
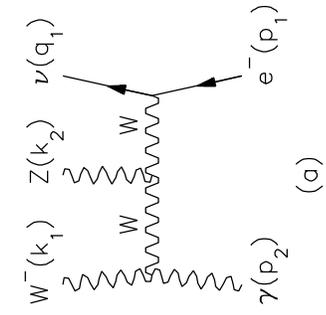
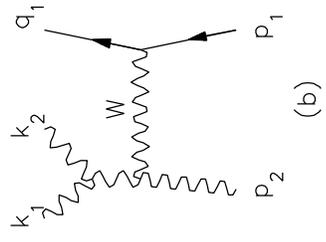
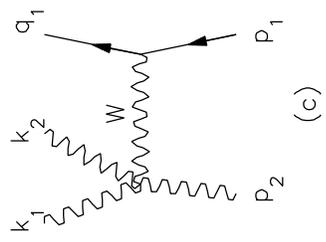
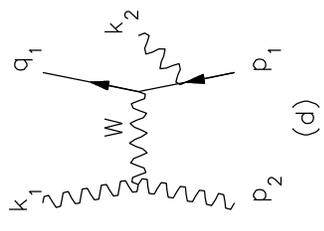
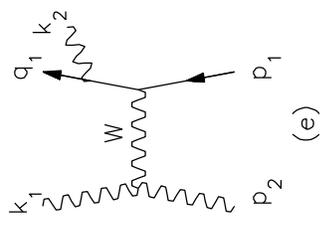
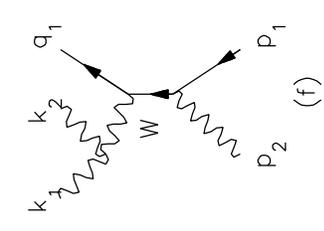
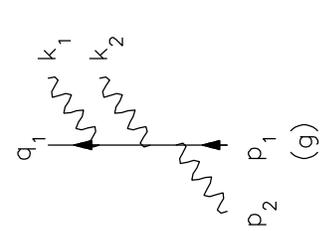
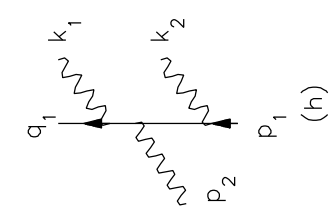
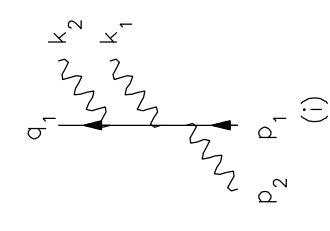
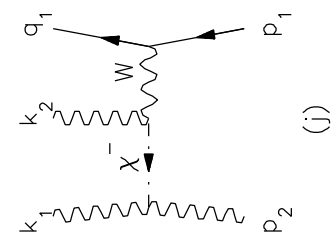
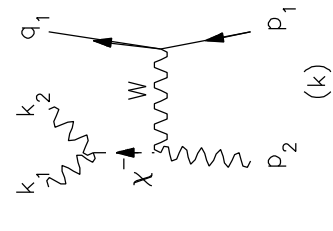
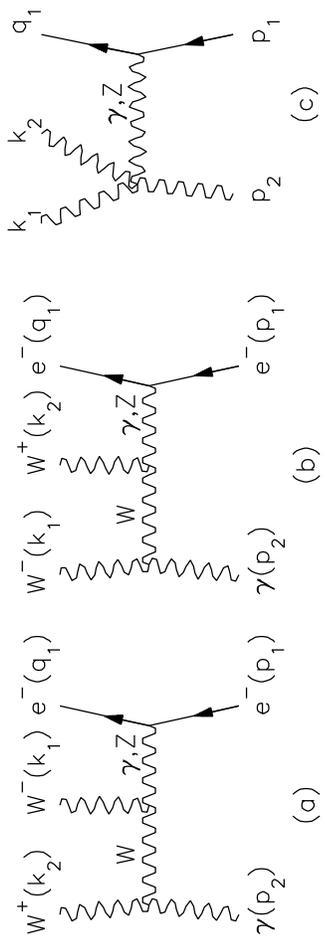
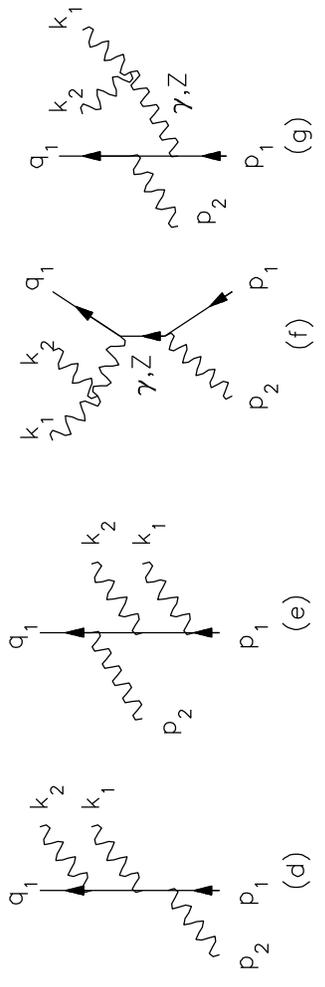
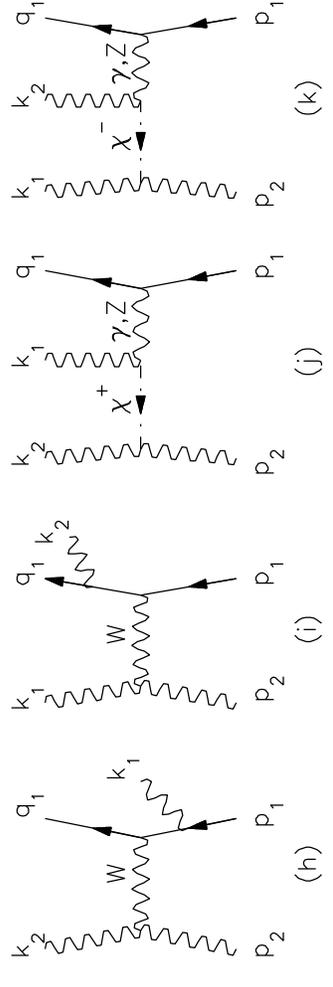
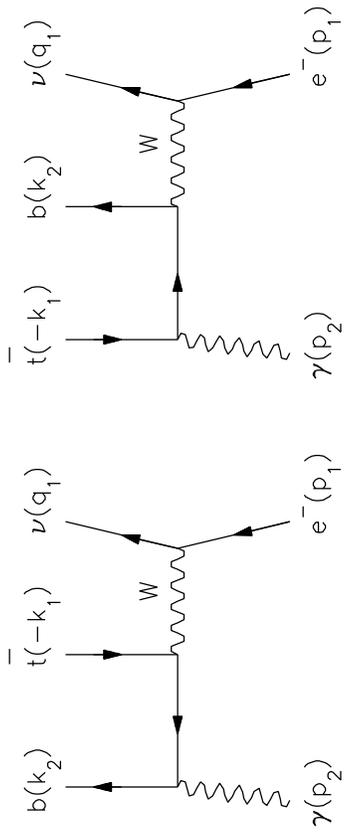
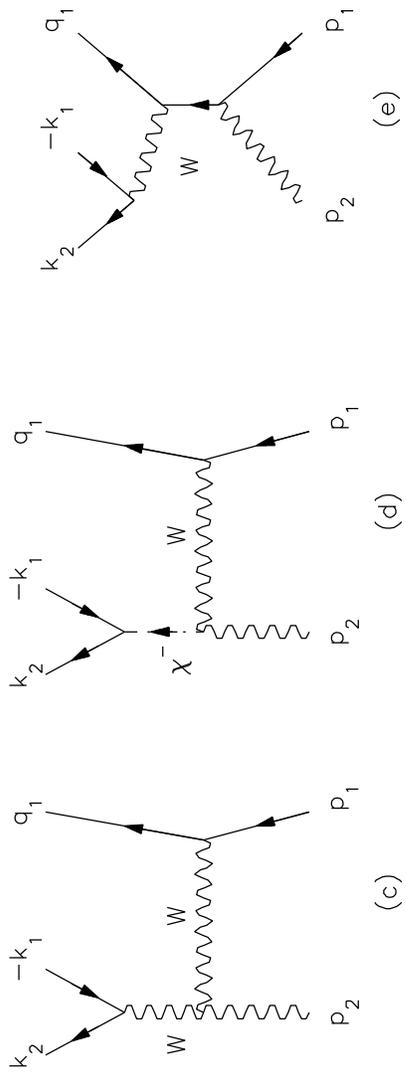
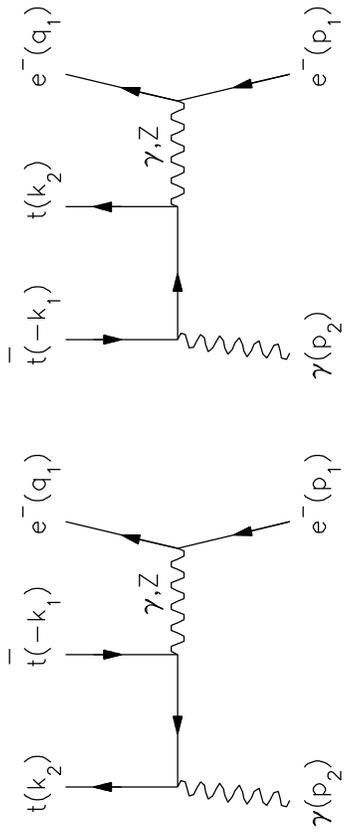
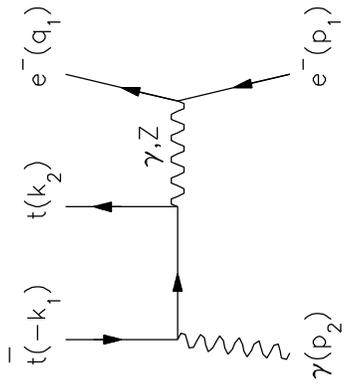
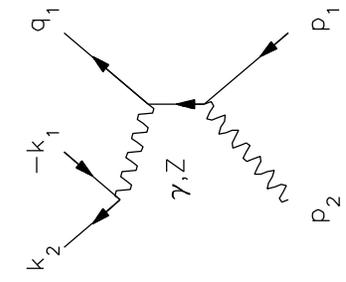
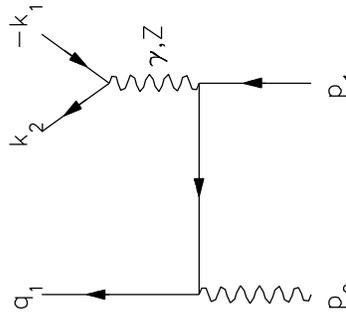
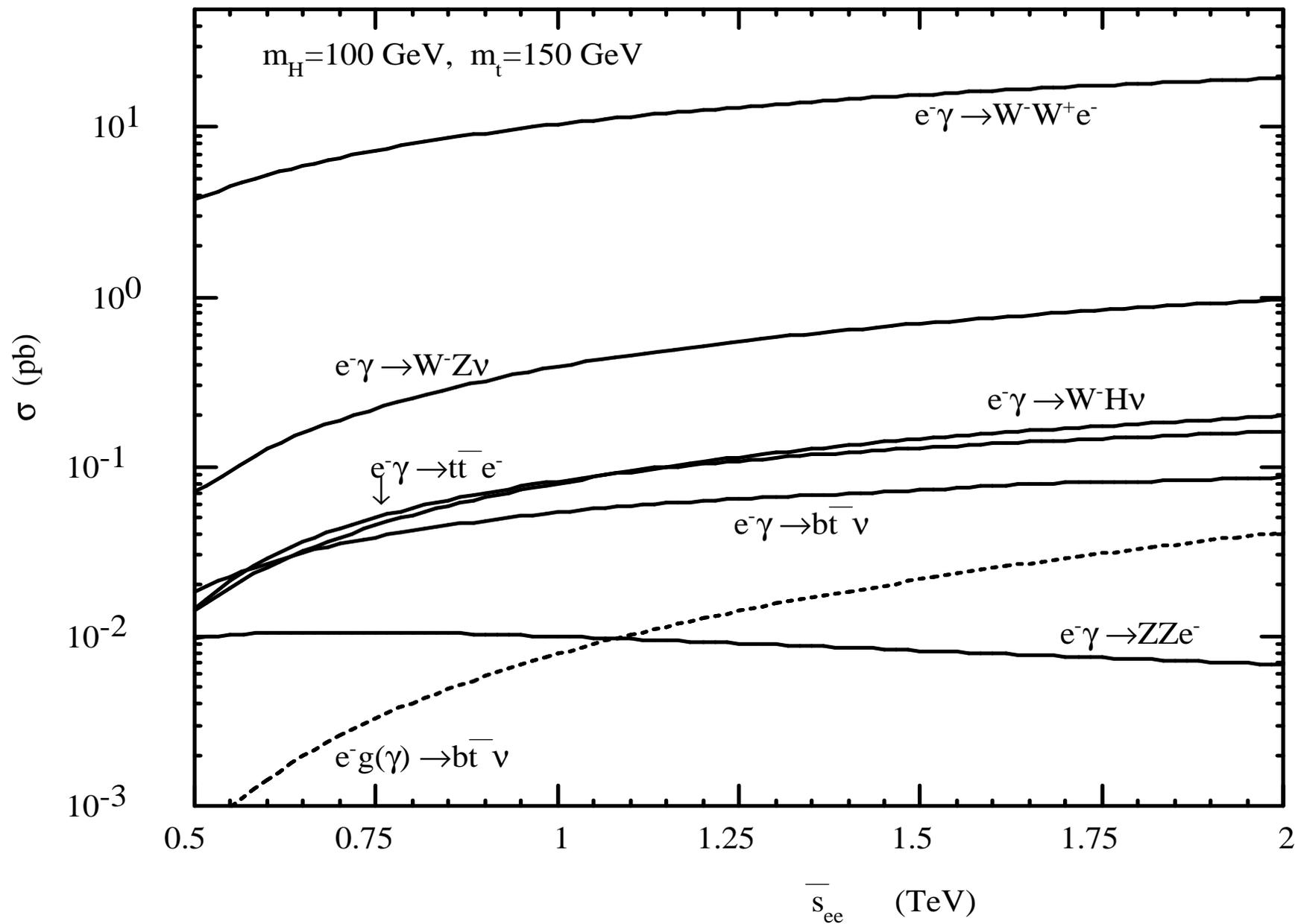
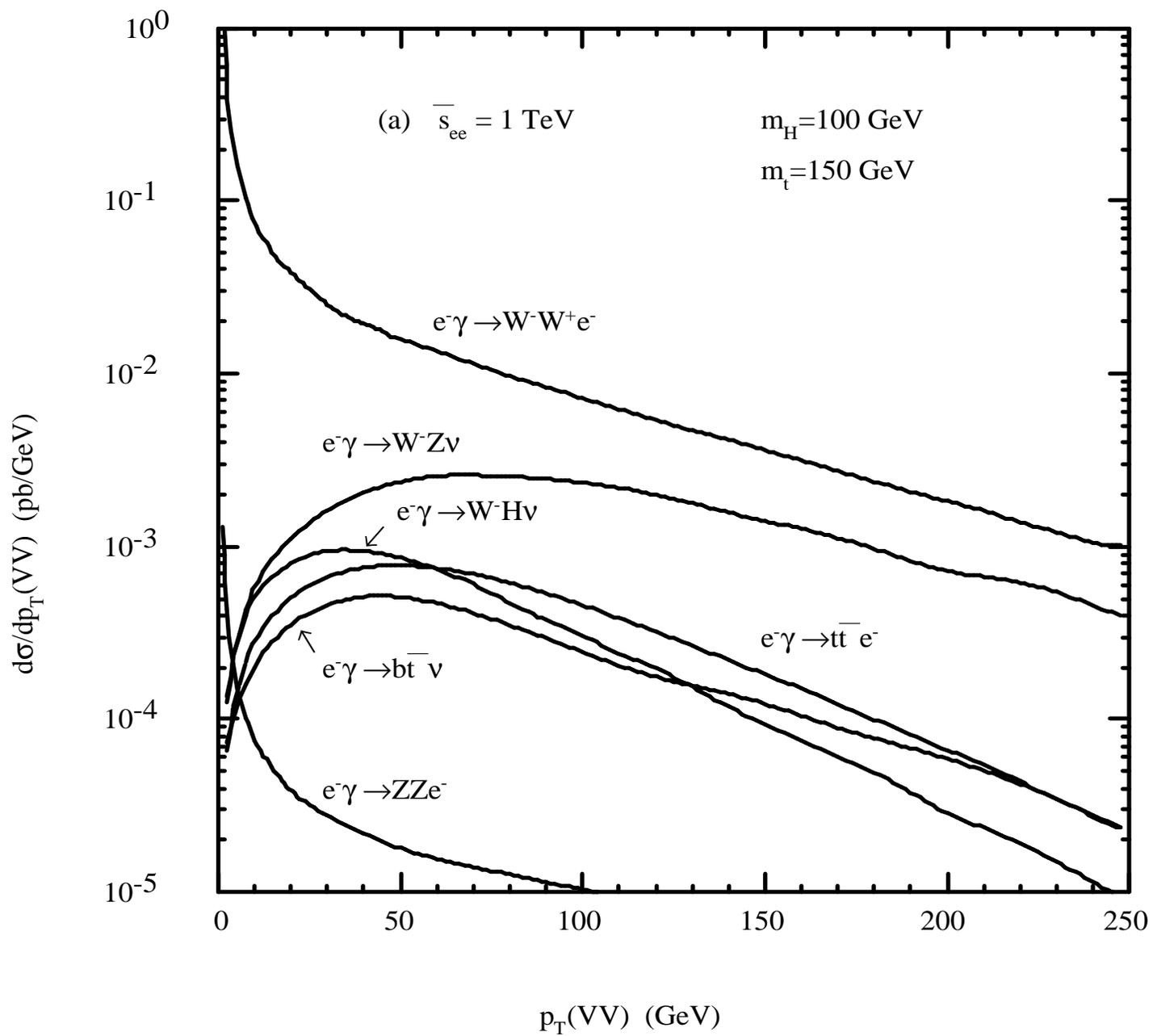
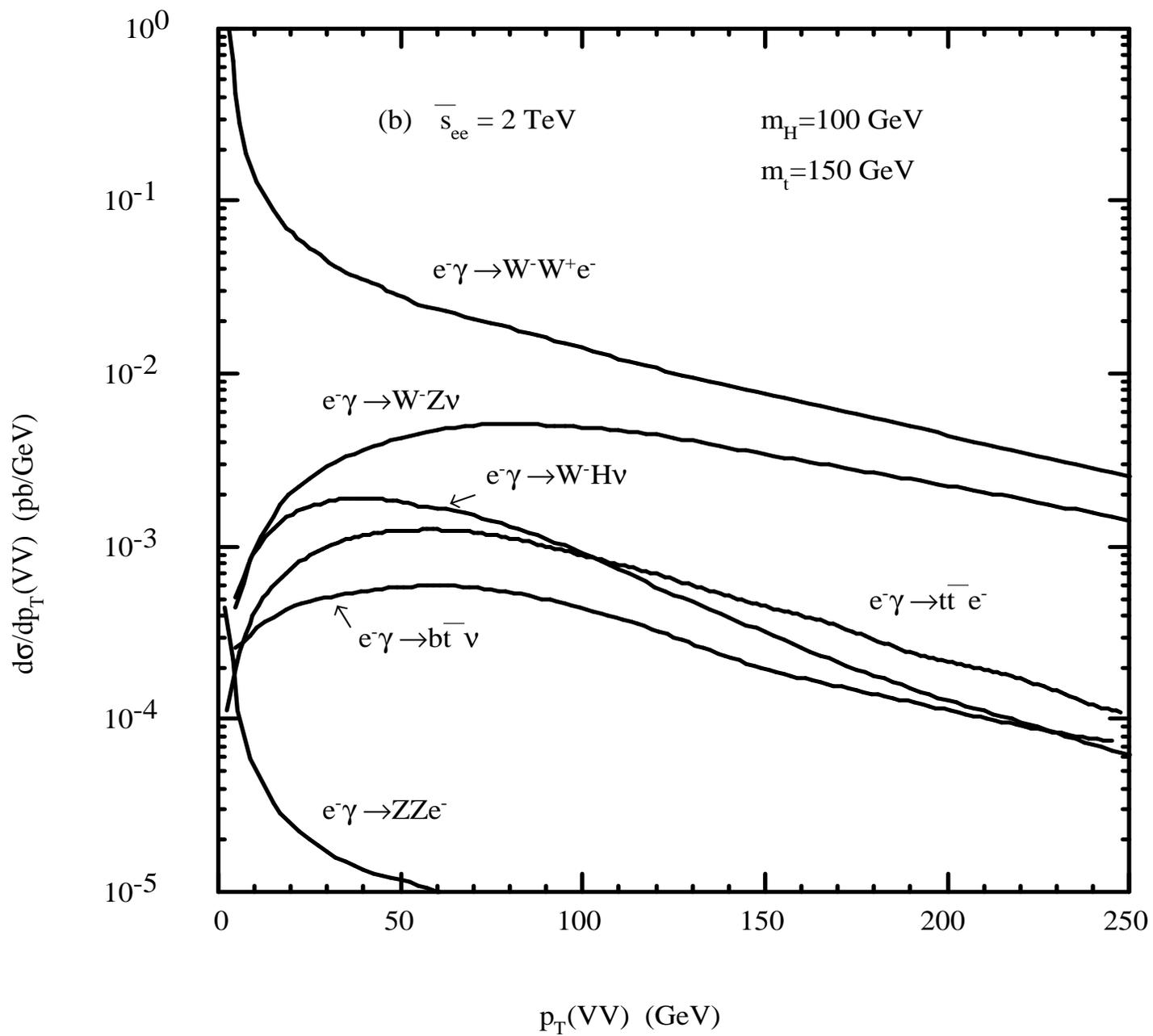
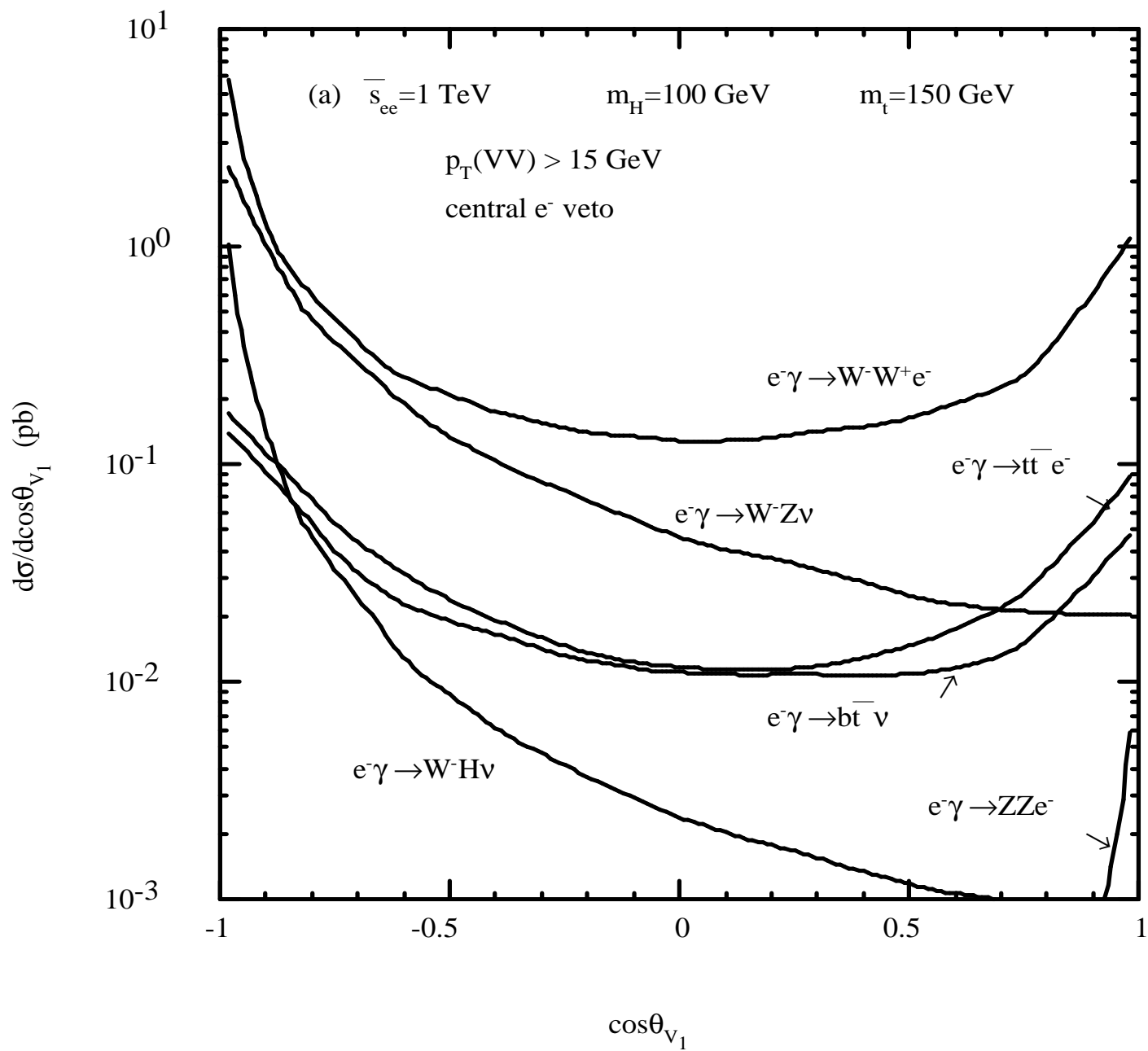
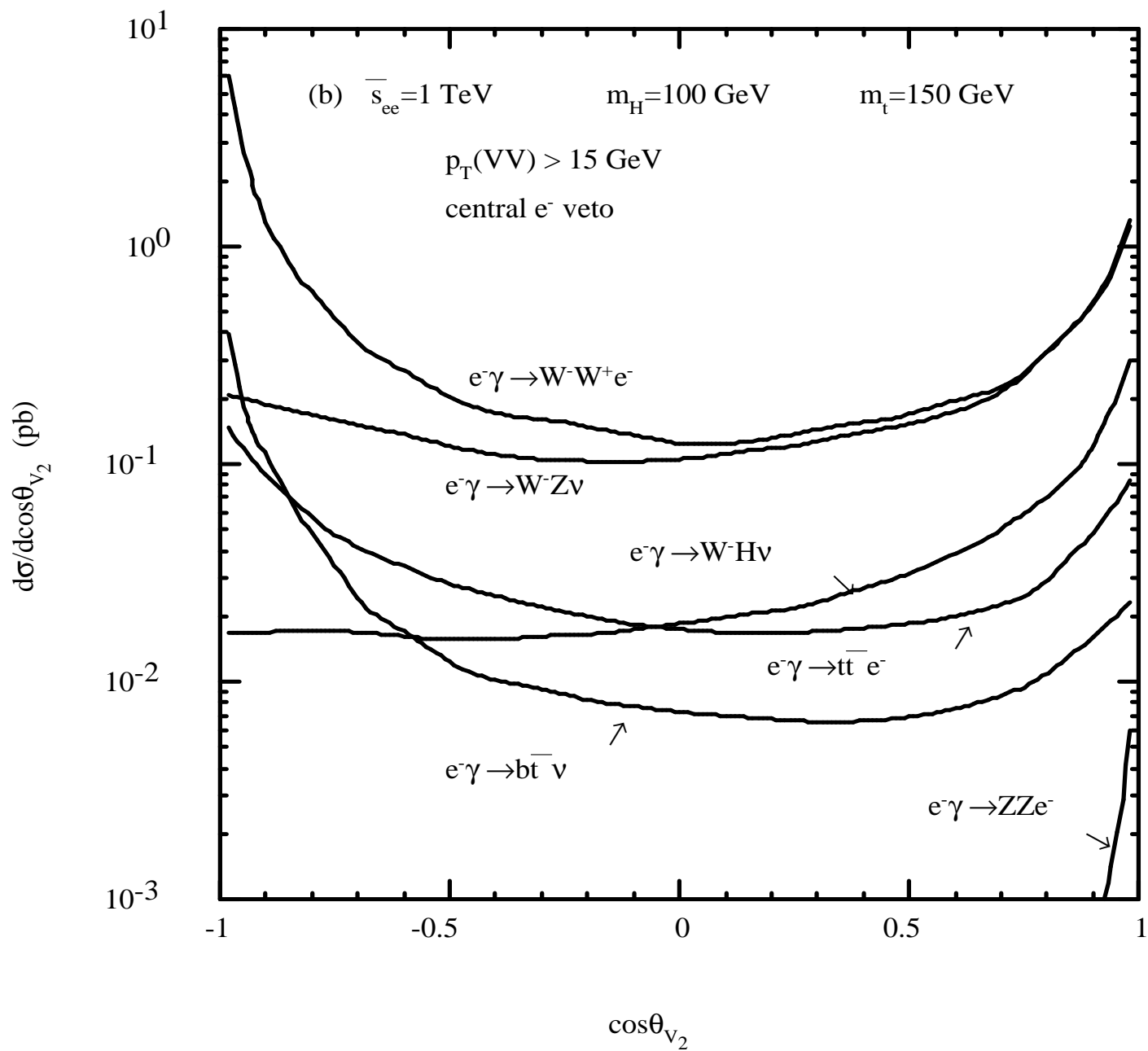
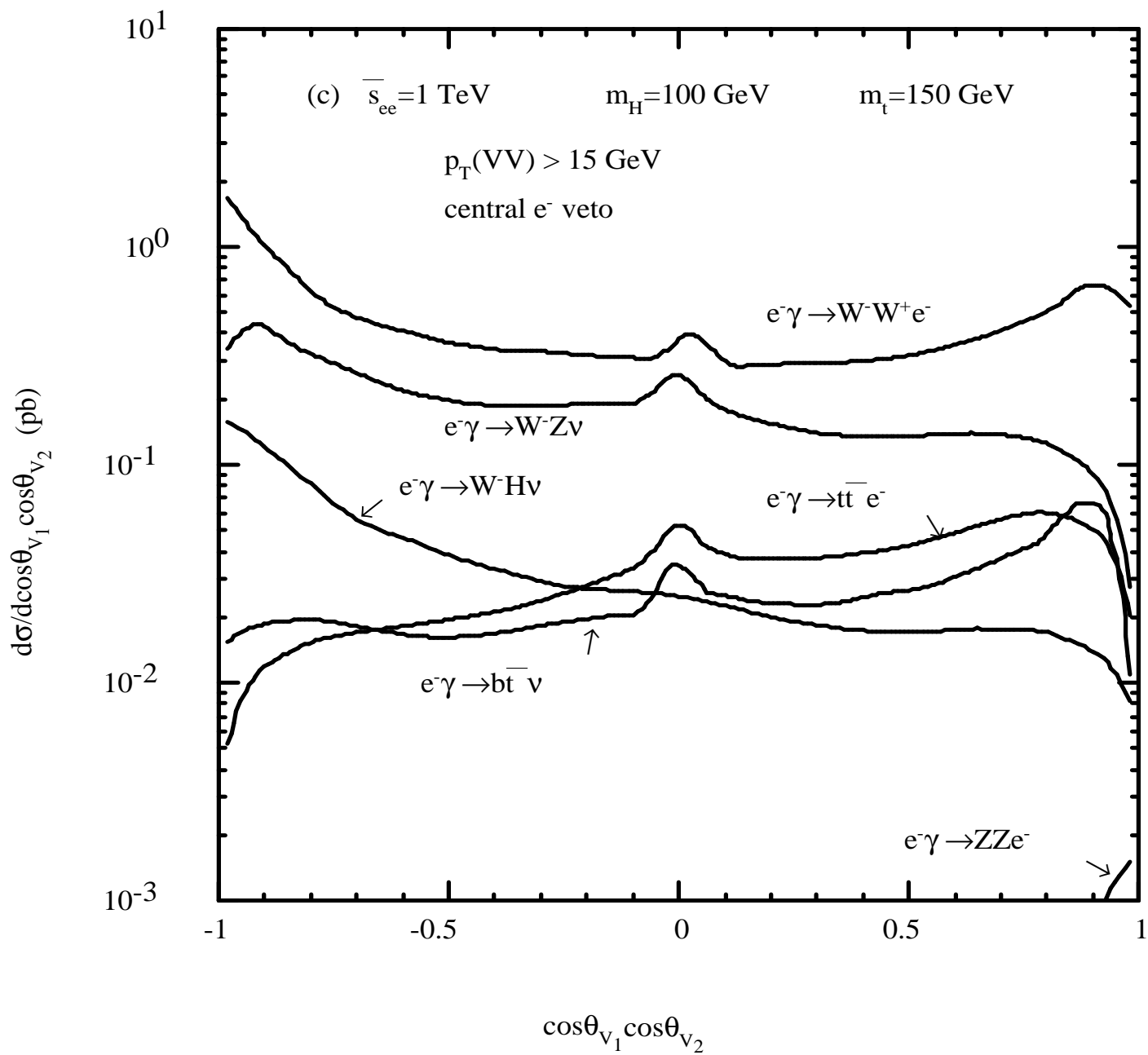
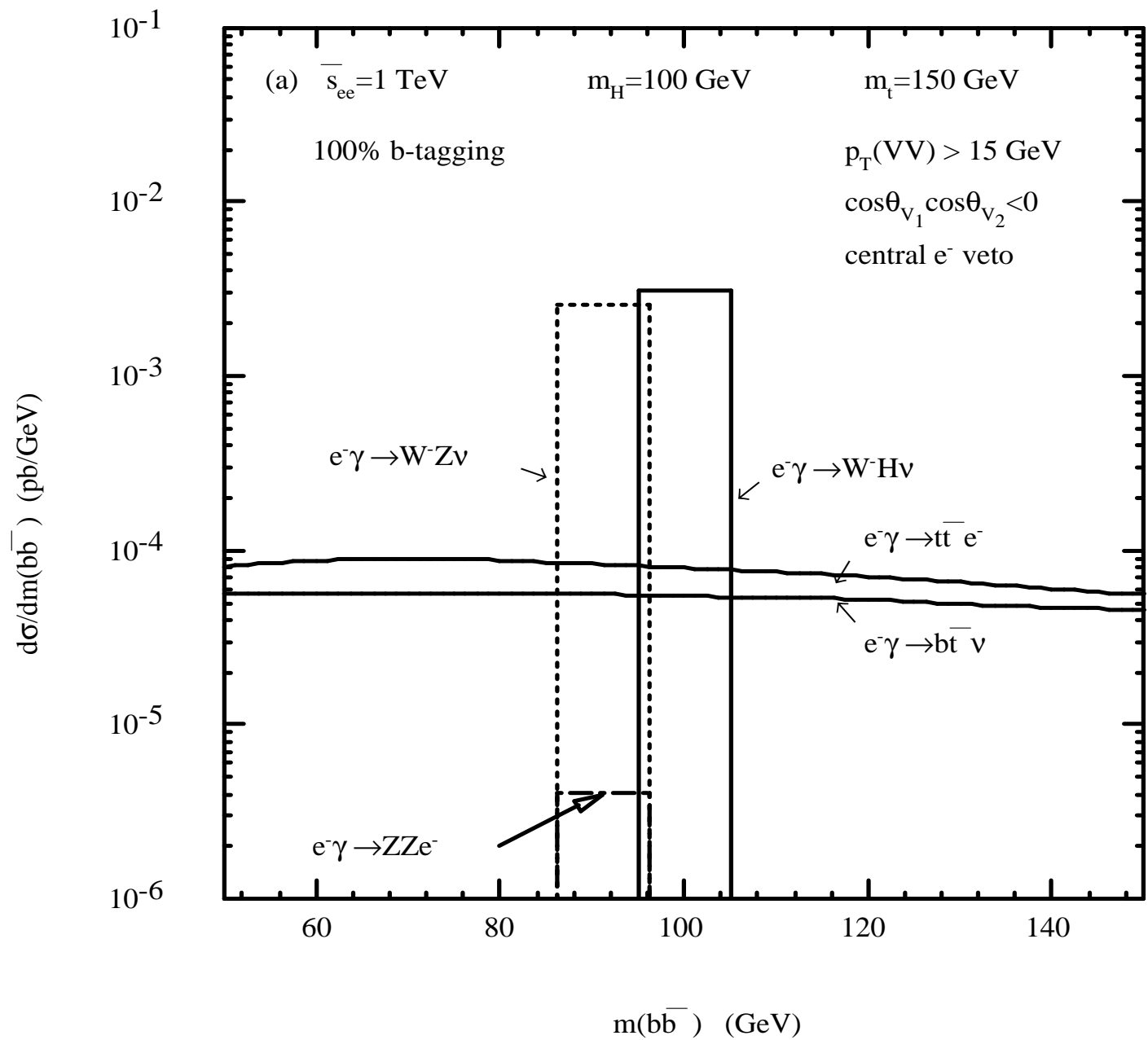
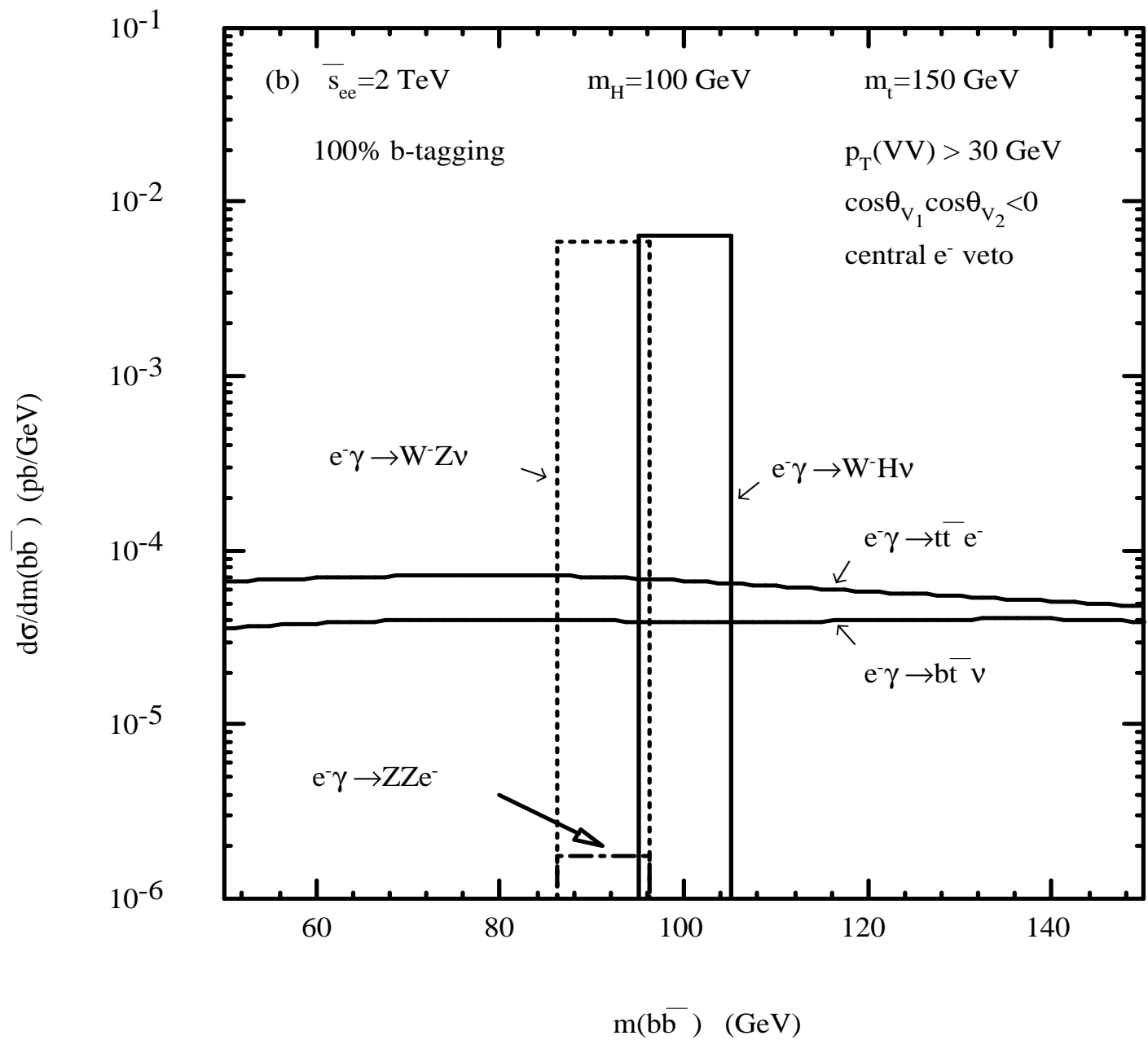
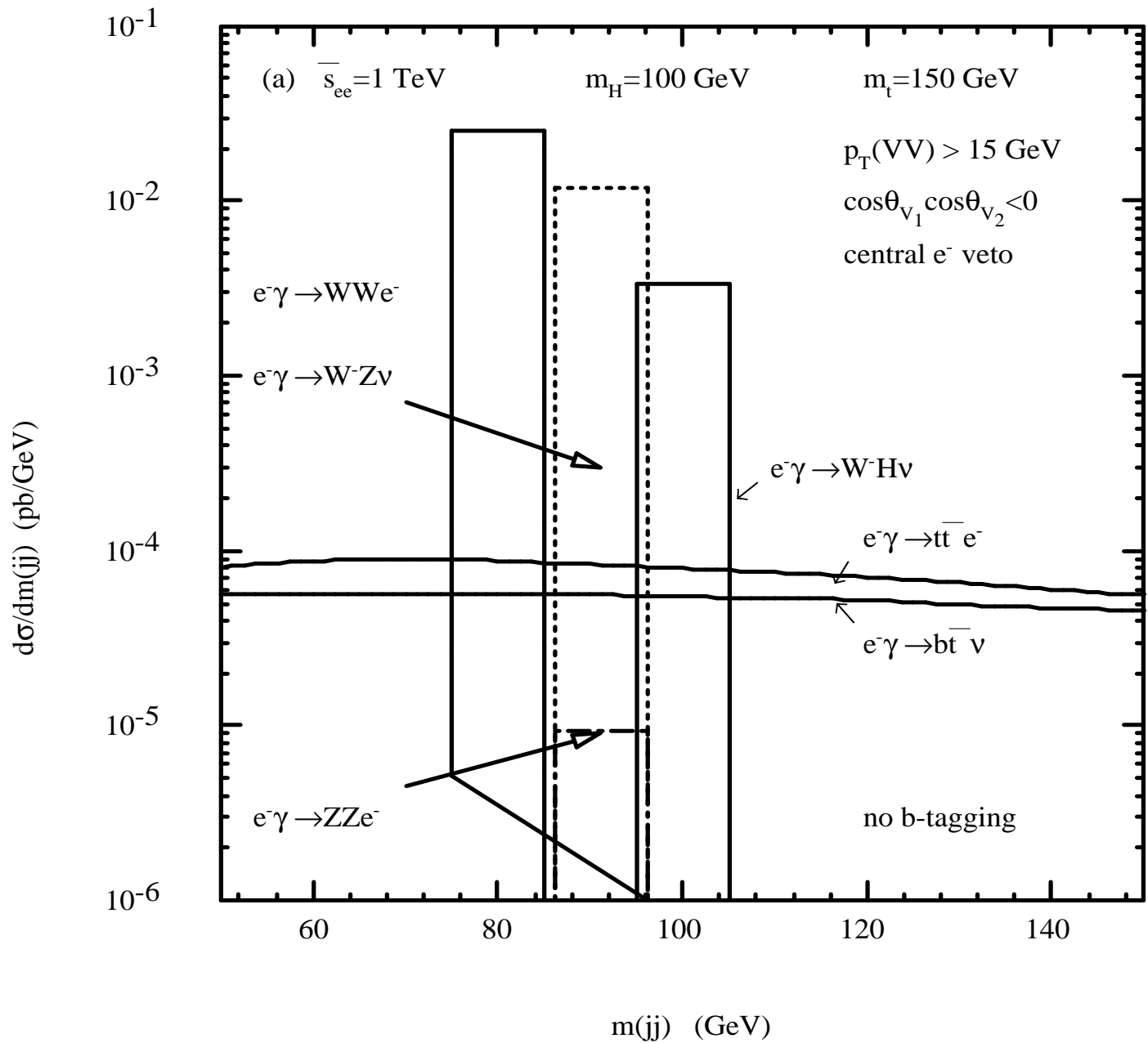
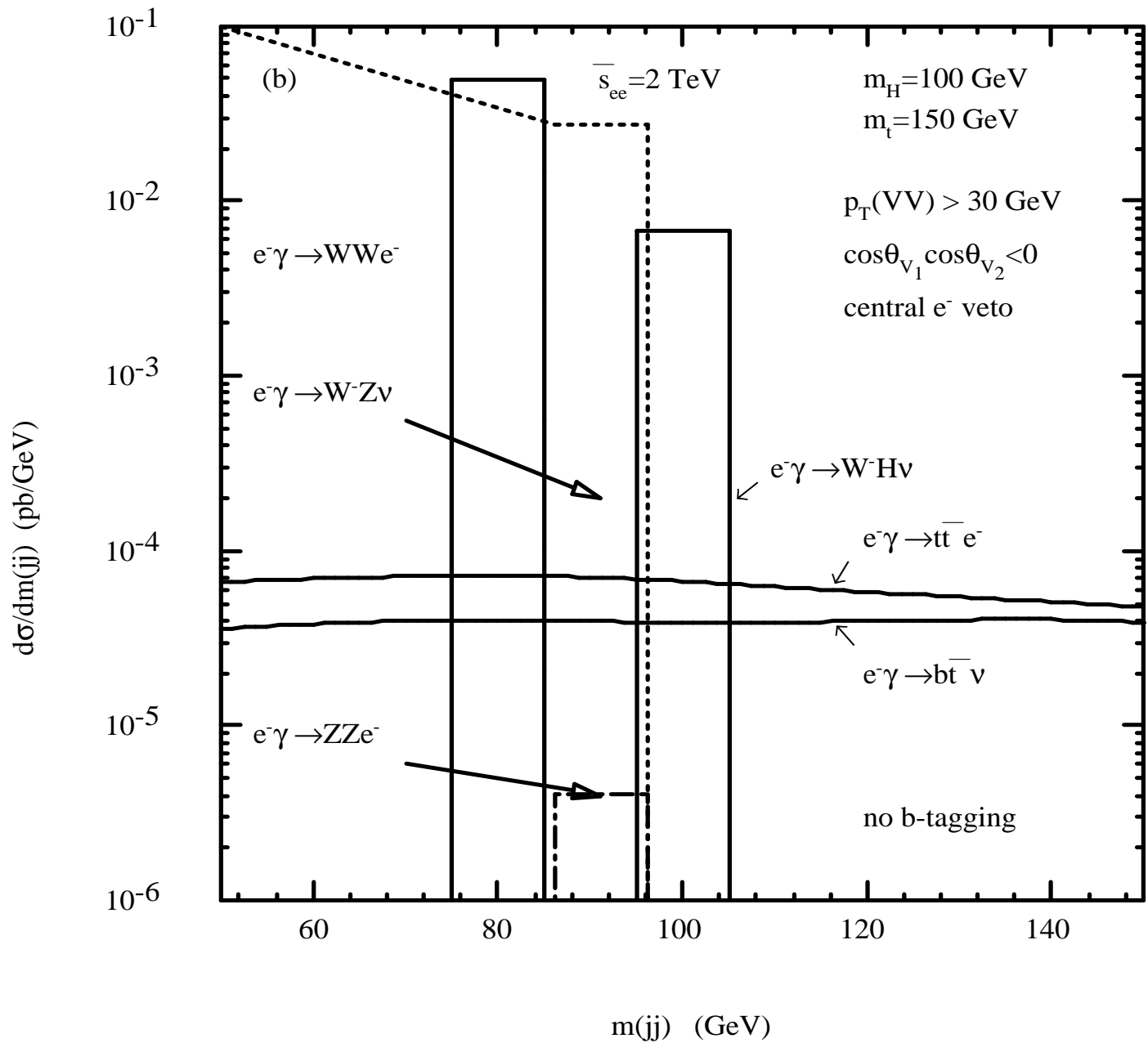
\end{document}